\documentclass[prd,aps,showpacs,nofootinbib,floats,letterpaper,floatfix,superscriptaddress,groupedaddress,eqsecnum,twocolumn]{revtex4-1}
\usepackage{graphicx,epsf, epsfig, amssymb}
\usepackage{amsfonts,amsmath,amssymb,amsthm}
\usepackage{bm}
\usepackage{hyperref}
\usepackage{latexsym}
\usepackage{rotating}
\usepackage[dvips]{color}
\usepackage{longtable}

\def\ii{{\rm i}}
\newcommand{\pa}{\partial}
\newcommand{\nn}{\nonumber}

\begin{document}

\title{Gravitational waves from extreme mass-ratio inspirals\\ in Dynamical Chern-Simons gravity}

\author{Paolo Pani} \email{paolo.pani@ist.utl.pt} \affiliation{CENTRA, Departamento de F\'{\i}sica, 
Instituto Superior T\'ecnico, Universidade T\'ecnica de Lisboa,
Av.~Rovisco Pais 1, 1049 Lisboa, Portugal} 

\author{Vitor Cardoso} \email{vitor.cardoso@ist.utl.pt}
\affiliation{CENTRA, Departamento de F\'{\i}sica, 
Instituto Superior T\'ecnico, Universidade T\'ecnica de Lisboa,
Av.~Rovisco Pais 1, 1049 Lisboa, Portugal} 
\affiliation{Department of Physics and Astronomy, The University of
 Mississippi, University, MS 38677-1848, USA}

\author{Leonardo Gualtieri} \email{leonardo.gualtieri@roma1.infn.it} \affiliation{Dipartimento di Fisica, 
``Sapienza''  Universit\`a di Roma and Sezione INFN Roma1, P.A. Moro 5, 00185, Roma, Italy}

\begin{abstract} 
  Dynamical Chern-Simons gravity is an interesting extension of General Relativity, which finds its way in many
  different contexts, including string theory, cosmological settings and loop quantum gravity.  In this theory, the
  gravitational field is coupled to a scalar field by a parity-violating term, which gives rise to characteristic
  signatures.  Here we investigate how Chern-Simons gravity would affect the quasi-circular inspiralling of a small,
  stellar-mass object into a large non-rotating supermassive black hole, and the accompanying emission of gravitational
  and scalar waves.  We find the relevant equations describing the perturbation induced by the small object, and we
  solve them through the use of Green's function techniques. Our results show that for a wide range of coupling
  parameters, the Chern-Simons coupling gives rise to an {\it increase} in total energy flux, which translates into a
  fewer number of gravitational-wave cycles over a certain bandwidth. For space-based gravitational-wave detectors such
  as LISA, this effect can be used to constrain the coupling parameter effectively.
\end{abstract}


\maketitle
\section{Introduction}\label{intro}
The inspirals of stellar-mass compact objects, typically black holes (BHs) or neutron stars, into supermassive BHs at
the galactic centers are among the most promising sources for the space-based gravitational wave detector LISA
\cite{AmaroSeoane:2007aw}.  These processes are known as Extreme Mass-Ratio Inspirals (EMRIs) and allow for stringent
tests of general relativity to be done, for several reasons \cite{Schutz:2009tz,2010GWN.....4....3S}. First of all, they
emit $10^5$ cycles or more of gravitational radiation in the LISA band, during the timespan of the LISA mission
($\sim1\!-\!5$ years). Furthermore, this signal is emitted when the stellar mass object is close to the horizon of the
supermassive BH, thus encoding the features of the strong-field BH spacetime and of the strong-curvature regime of
general relativity~\cite{Ryan:1995wh}. Finally, EMRIs are expected to be relatively clean systems, with negligible
perturbations from surrounding matter. Therefore, the detection of the signal from an EMRI would allow us to test the
strong-field regime of gravity, where possible deviations from general relativity may show up. Of course, this would be
possible only if we understand how alternative theories of gravity would affect the EMRI signal.

In this paper, we discuss the imprint on EMRI signals of dynamical Chern Simons (DCS) gravity, an extension of
general relativity \cite{Deser:1982vy,Lue:1998mq,Jackiw:2003pm,Smith:2007jm} in which the Einstein-Hilbert action is
modified by adding a parity-violating Chern-Simons (CS) term, that couples gravity to a scalar field. Among other
proposed alternative theories, there are strong motivations to consider DCS gravity. Indeed, this correction arises in
many versions of string theory \cite{Polchinski:1998rr} and in loop quantum gravity
\cite{Ashtekar:1988sw,Taveras:2008yf,Mercuri:2009zt}. Furthermore, CS gravity can be recovered as a truncation of low
energy effective string models~\cite{Smith:2007jm,Adak:2008yg} and could also explain several problems in
cosmology~\cite{Weinberg:2008mc,GarciaBellido:2003wd,Alexander:2004xd,Alexander:2004us,Konno:2008np} (see
Ref.~\cite{Alexander:2009tp} for a recent review).

DCS gravity would affect the EMRI signal in several ways. First of all, it modifies the spacetime metric of rotating
BHs~\cite{Yunes:2009hc} (though the metric of spherical, stationary BHs is not affected). These modifications induce
deviations in the motion of the stellar mass object, which affect the emitted gravitational signal.  This effect has
been studied in \cite{Sopuerta:2009iy}, where the deviation of EMRI orbits due to spacetime metric modification has been
evaluated. In addition, in this theory the dynamical equations of gravity change, as the gravitational field is coupled
with a scalar field. This is the effect studied in this paper, in which we evaluate the change of the EMRI signal due to
the DCS modification of the dynamical equations of the gravitational field.

We shall focus on the simplest case of EMRI system: the inspiralling of a test particle around a Schwarzschild BH. Since
the Schwarzschild metric is a solution of DCS gravity, the orbital effects studied in \cite{Sopuerta:2009iy}
vanish. However, as we will show, DCS gravity could significantly affect the EMRI signal.  We generalize the equations
describing perturbations of Schwarzschild DCS BHs, derived previously by some of us \cite{Cardoso:2009pk,Molina:2010fb},
to include a source term describing an orbiting particle. Then, we solve these equations using a generalization of the
Green's function method, evaluating the emitted flux of gravitational energy and scalar energy. Finally, we determine
how the modified energy loss affects the emitted gravitational signal, by computing the change in the number of orbital
cycles.  Similar methods have been already successfully applied to study the effects of neutron star structure in EMRI
systems~\cite{Pons:2001xs}, to investigate the nature of the central massive object~\cite{Pani:2010em} and also to put
constraints on other alternative theories of gravity, like Brans-Dicke theory~\cite{Ohashi:1996uz}.

We find that, although the DCS coupling only mildly changes the total energy flux, the corrections accumulate during the
EMRI inspiral, producing a decrease $\delta{\cal N}$ in the number of orbital cycles which is potentially detectable by LISA. Our results
can be summarized by the following fit:
\begin{equation}
\delta{\cal N}\sim-26\zeta\sqrt{\frac{M_\odot}{m_2}}\exp\left\{-1.2\log_{10}^2\left[\frac{m_1}{m_\text{max}}\right]\right\}\,,\label{fit0}
\end{equation}
where $m_\text{max}\sim10^6 M_\odot$ (its precise definition is given in Eq.~\eqref{mmax}), $\zeta$ is a parameter
characterizing the DCS correction, and $m_1,m_2$ are the masses of the supermassive BH and of
the stellar mass object, respectively. The fit above is valid in a wide region of the parameter space, i.e.  $m_1\in[10^5,10^7]M_\odot$,
$m_2\in[1,10]M_\odot$ and $\zeta\le1$, with an error of at most a few percent. For larger values of $\zeta$, the fit (\ref{fit0}) still gives
an order-of-magnitude estimate of $\delta{\cal N}$: for $1\le\zeta\le20$, the error does not exceed $50\%$.

The plan of the paper is the following. In Section~\ref{dcsperteq} we discuss the equations describing perturbations of
the Schwarzschild metric in DCS gravity induced by a point-like particle on a circular orbit about the BH. In Section~
\ref{green} we describe the Green's function approach to solve the perturbation equations. In Section~\ref{flux} we
derive the energy flux associated to the gravitational and scalar radiation, and compute this flux solving numerically
the perturbation equations. In Section~\ref{dcycles} we determine how this energy flux affects the gravitational
signal. In Section~\ref{concl} we draw our conclusions. The derivation of the perturbation equations with source is
discussed in detail in Appendix \ref{ZRW}. Appendix~\ref{app:green} is devoted to describe a perturbative Green's
function approach, which is valid in the small coupling limit and may be potentially useful for future analytical
calculations. In Appendix~\ref{app:comparison} we compare different prescriptions to compute the number of orbital
cycles.
\section{DCS gravity and perturbation equations }\label{dcsperteq}
The action of DCS gravity reads \cite{Yunes:2009hc} (we use geometrical units $c=G=1$)
\begin{eqnarray}
&&S=\frac{1}{16\pi}\int d^4x\sqrt{-g}R+\frac{\alpha}{4}\int d^4x\sqrt{-g}
\vartheta\,^*RR\nonumber\\
&&-\frac{\beta}{2}\int d^4x\sqrt{-g}\left[ 
g^{ab}\nabla_a\vartheta\nabla_b\vartheta+V(\vartheta)\right]+S_{\rm mat}\,,\label{action}
\end{eqnarray}
where $\vartheta$ is the scalar field and 
\begin{equation}
^*RR=R_{abcd}{}^*R^{bacd}=\frac{1}{2}R_{abcd}\epsilon^{baef}R^{cd}_{~~ef}\,.
\end{equation}
Neglecting the scalar potential $V(\vartheta)$, the equations of motion are
\begin{eqnarray}
R_{ab}&=&-16\pi \alpha C_{ab}+8\pi\left(T_{ab}-\frac{1}{2}g_{ab}T\right)\label{eqE1}\\
\Box\vartheta&=&-\frac{\alpha}{4\beta}\,^*RR \label{eqE2}
\end{eqnarray}
where the stress-energy tensor accounts for the matter and the scalar field contributions, 
$T_{ab}=T_{ab}^\text{mat}+T_{ab}^{\vartheta}$, with
\begin{equation}
T_{ab}^{\vartheta}=\beta\left(\vartheta_{;a}\vartheta_{;b}-\frac{1}{2}\vartheta_{;c}\vartheta^{;c}\right)\,,
\end{equation}
and
\begin{equation} 
C^{ab}=\vartheta_{;c}\epsilon^{cde(a}\nabla_eR^{b)}_{~~d}
+\vartheta_{;dc}\,^*R^{d(ab)c}\,.\label{Ctensor}
\end{equation} 
Since in any spherically symmetric background $^*RR\equiv0$ and $C^{ab}\equiv0$, spherically symmetric solutions of
general relativity, like the Schwarzschild metric
\begin{equation}
ds^2=-fdt^2+f^{-1}dr^2+r^2d\Omega^2\,,\label{schwmet}
\end{equation}
are solutions of DCS gravity~\cite{Alexander:2009tp}. Here, $M$ is the ADM mass of the spacetime and
\begin{equation} 
f=1-2M/r\,.
\end{equation} 

A remarkable feature of EMRIs is that they can be described with great accuracy within a perturbative approach and in an
adiabatic approximation. Indeed, during most of the inspiral the stellar-mass object can be considered as a test
particle moving in a single massive BH background, the timescale for merger being much longer than a single orbital
period. Hence, at each instant, we consider that the particle follows a geodesic of the BH spacetime and the geodesic
parameters, i.e. the orbital energy and angular momentum of the particle, would change \emph{adiabatically}: they can be
computed by solving the linearized Einstein's equations for geodesic motion. In this way one finds the inspiralling
orbit and the corresponding gravitational waveform, as explained below. This procedure takes into account the main
effect of the back-reaction (the so-called ``non-conservative part of the self-force''). A more detailed analysis, which
would also consider the ``conservative part of the self-force'' (see \cite{Barack:2007tm,Detweiler:2008ft} for the case
of non-rotating BHs and \cite{Poisson:2003nc,Barack:2009ux} for a review), is beyond the scope of this work.

Remarkably, due to Eq.~\eqref{eqE2}, the tensor~\eqref{Ctensor} satisfies $\nabla^a
C_{ab}=8\pi\nabla^aT_{ab}^\vartheta$.  Using the latter equation and the Bianchi identities, the matter stress-energy
tensor is conserved also in DCS gravity, i.e. $\nabla^{a}T_{ab}^\text{mat}=0$. It follows that point-like particles
travel on geodesics, exactly as in general relativity. Furthermore, we assume that the point-particle is non-spinning,
neglecting the effects of spin-orbit interactions.  These effects are proportional to the mass-ratio and they can be
safely neglected in the study of EMRI systems \cite{Tanaka:1996ht}.

We shall study the perturbations of a static, spherically symmetric BH of mass $M$, due to a non-spinning point-like particle of mass $\mu$ on a
circular orbit around the BH.  As discussed in \cite{Molina:2010fb}, the Schwarzschild metric~\eqref{schwmet} with a
vanishing scalar field is the only static, spherically symmetric BH solution in DCS gravity. In this background, we
expand the gravitational and scalar perturbations induced by a point-like particle in tensor spherical harmonics,
building the Zerilli and Regge-Wheeler functions, $Z^{\ell m}(r)$, $Q^{\ell m}(r)$, and the scalar field function
$\Theta^{\ell m}(r)$. In the frequency domain, the perturbation equations read (see Appendix~\ref{ZRW} for details):
\begin{eqnarray}
\left[\frac{d^2}{d r_*^2}\!+\!\omega^2\!-\!V_{RW}(r)\right]\!Q^{\ell m}(r)&=&T_{RW}(r)\Theta^{\ell
 m}(r)\!+\!S^{\ell m}_{RW}(r)\nonumber\\
&&\label{eqRW}\\
\left[\frac{d^2}{d r_*^2}\!+\!\omega^2\!-\!V_{S}(r)\right]\!\Theta^{\ell m}(r)&=&T_S(r)Q^{\ell m}(r)+S^{\ell m}_{S}(r)\nonumber \\ 
\label{eqScalar}\\
\left[\frac{d^2}{d r_*^2}\!+\!\omega^2\!-\!V_{Z}(r)\right]Z^{\ell m}(r)&=&S^{\ell m}_{Z}(r)\,.\label{eqZer}
\end{eqnarray}
In the equations above $r_*$ is the tortoise coordinate defined by $dr/dr_*=f$, and the potentials read
\begin{eqnarray}
&&V_{RW}(r)=f\left(\frac{\ell(\ell+1)}{r^2}-\frac{6M}{r^3}\right)\label{potentialRW}\\
&&T_{RW}(r)=f\frac{96i\pi M\omega \alpha}{r^5}\\
&&V_{S}(r)=f\left(\frac{\ell(\ell+1)}{r^2}\left[1+\frac{576\pi
M^2\alpha^2}{r^6\beta}\right]+\frac{2M}{r^3}\right)\label{potentialScalar}\\
&&T_S(r)=-f\frac{(\ell+2)!}{(\ell-2)!}\frac{6Mi\alpha}{r^5\beta\omega} \\
&&V_{Z}(r)=\frac{f}{r^2\Lambda^2}\Bigg[2\lambda^2\!\left(\lambda\!+\!1\!+\!\frac{3M}{r}\right)
\!+\!\frac{18M^2}{r^2}\left(\lambda\!+\!\frac{M}{r}\right)\Bigg]\,,\nonumber\\
&&\label{potentialZer}
\end{eqnarray}
where $\lambda=(\ell+2)(\ell-1)/2$ and $\Lambda=\lambda+3M/r$. The source terms are given in Appendix~\ref{ZRW}:
$S^{\ell m}_\text{RW}$ and $S^{\ell m}_Z$ (cf. Eqs.~\eqref{sourceRW} and~\eqref{sourceZer}) are the same as those
computed in general relativity~\cite{Martel:2003jj}, whereas $S^{\ell m}_S$ (cf. Eq.~\eqref{sourceScalar}) is
proportional to $\alpha$ and it is a novel term introduced by the CS coupling.

We stress that Eqs.~\eqref{eqRW}~and~\eqref{eqScalar} are coupled through the CS coupling $\alpha$. In the sourceless
case (i.e., no exterior matter), they reduce to those in Refs.~\cite{Cardoso:2009pk,Molina:2010fb} with $Q^{\ell
  m}=i\omega \Psi^{\ell m}$ (notice the sign difference due to our definition \eqref{defRW}). In the general relativity
limit $\alpha=0$, they decouple into the Regge-Wheeler equation with sources \cite{Davis:1972dm,Martel:2003jj} plus
Klein-Gordon equation without source. Interestingly, in the DCS Schwarzschild background the polar (even parity)
gravitational sector, Eq.~\eqref{eqZer}, decouples from the scalar sector. Thus the CS coupling does not affect the
Zerilli equation, which simply reads as in general relativity \cite{Martel:2003jj}.

To conclude this Section, we remark that if we rescale the scalar field, in order to express its kinetic term in a
canonical form
\begin{equation}
\theta\rightarrow\frac{\theta}{\sqrt{\beta}}\,,\label{rescaling}
\end{equation}
then the DCS action takes the form
\begin{eqnarray}
&&S=\frac{1}{16\pi}\int d^4x\sqrt{-g}R+\frac{\sqrt{\xi}}{16\sqrt{\pi}}\int d^4x\sqrt{-g}
\vartheta\,^*RR\nonumber\\
&&-\frac{1}{2}\int d^4x\sqrt{-g}\left[
g^{ab}\nabla_a\vartheta\nabla_b\vartheta\!+\!\tilde V(\vartheta)\right]\!+\!S_{\rm mat}\,,\label{actionresc}
\end{eqnarray}
where, following Ref.~\cite{Yunes:2009hc}, we have defined $\xi\equiv{16\pi\alpha^2}/{\beta}$.  Therefore, we expect
that all physical observables depend on the parameter $\xi$ (see also the discussion in \cite{Yunes:2009hc}), or
equivalently on the dimensionless parameter
\begin{equation}
\zeta=\frac{16\pi\alpha^2}{\beta M^4}=\frac{\xi}{M^4}\,,\label{zetadef}
\end{equation}
where $M$ is a quantity with the dimensions of mass (i.e. of length) associated to the system under consideration (in
our case, the mass of the supermassive BH). This is indeed the case, as we shall show, for the motion of a test particle
around a static, spherically symmetric BH. This procedure, of rescaling the scalar field to have a canonical kinetic
term in order to get rid of redundant parameters, is well known in the context of scalar-tensor theories (see
\cite{2003sttg.book.....F} and references therein). However, here we shall follow the formulation of
Ref.~\cite{Yunes:2009hc}, where $\alpha$ and $\beta$ are kept as independent parameters.

%


\section{Green's function approach}\label{green}
In order to compute the gravitational-wave emission of a particle in geodesic motion around a spherically symmetric BH
in DCS gravity, we shall solve the equations~\eqref{eqRW}-\eqref{eqZer} by extending the standard Green's function
techniques. As explained below, we work out the basic equations for general orbits, and then we specialize to circular
motion. Our results can be easily generalized to eccentric orbits. In Appendix~\ref{app:green} we develop a perturbative
approach, which is valid in the small-coupling limit, and we compare it with the general method discussed in this
section (the agreement is very good for small coupling).
\subsection{Even sector}\label{greeneven}
We start by considering the Zerilli equation~\eqref{eqZer}, which is not modified in DCS gravity.
One considers two solutions $Z^{\ell m}_\pm$ of the associate homogeneous equation
\begin{equation}
\left[\frac{d^2}{d r_*^2}+\omega^2-V_Z\right]Z_\pm=0\,,\label{zerhomeq}
\end{equation}
(hereafter, we leave implicit the $\ell,m$ indices) such that
\begin{equation}
Z_\pm\rightarrow e^{\pm\ii\omega r_*}\,,~~~~~r_*\rightarrow\pm\infty\,.\label{BCpolar}
\end{equation}
Then the general solution reads
\begin{equation} 
Z(r)=\frac{1}{W_Z}\left[Z_+(r)\int_{-\infty}^r\!\!dr_*{Z_-S_Z}\!+\!Z_-(r)\int^{+\infty}_r\!\!dr_*{
Z_+S_Z}
\right]\,,\label{zerfunc}
\end{equation} 
where $W_Z\equiv f(Z_-Z_+'-Z_+Z_-')$ is the Wronskian and the prime denotes derivative with respect to the Schwarzschild
radial coordinate, $r$.  At infinity and at the horizon, where the energy flux is computed in term of $Z$, we get
\begin{equation}
Z(r_*\rightarrow\pm\infty)=\frac{e^{\pm\ii\omega r_*}}{W_Z}\int_{-\infty}^\infty dr_*{Z_\mp S_Z}\,.\label{zerfuncinf}
\end{equation}
As we show in Appendix~\ref{ZRW}, for a circular orbit at $r=\bar r$ the source term has the form $S_Z\sim\delta(r-\bar
r)$. In this case to compute the integral (\ref{zerfuncinf}) it is sufficient to evaluate the integrand at $r=\bar r$.
\subsection{Odd sector}
Let us consider the modified Regge-Wheeler equation~\eqref{eqRW}, coupled with the scalar equation~\eqref{eqScalar}:
\begin{eqnarray}
\left[\frac{d^2}{d r_*^2}+\omega^2-V_{RW}\right]Q&=&S_{RW}+\frac{96\ii\pi M\omega
f}{r^5}\alpha\Theta\,,\label{rweq}\\
\left[\frac{d^2}{d r_*^2}+\omega^2-V_S\right]
\Theta&=&S_S
-f\frac{(\ell+2)!}{(\ell-2)!}\frac{6\ii M\alpha}{\omega r^5\beta}Q\,.\nonumber\\
\label{sceq}
\end{eqnarray}
Introducing $P^{\ell m}=dQ^{\ell m}/dr_*$ and $\Phi^{\ell m}=d\Theta^{\ell m}/dr_*$, we can write the equations above as
a first order system
\begin{equation}
 \frac{d\mathbf{\Psi}}{d r_*}+V\mathbf{\Psi}=\mathbf{S}\,,\label{system}
\end{equation}
where $\mathbf{\Psi}=(Q,\Theta,P,\Phi)^T$ and $\mathbf{S}=(0,0,S_{RW},S_{S})^T$ are four dimensional vectors and $V$ reads
\begin{equation}
 V= \begin{pmatrix} 0&0&-1&0\\ 0&0&0&-1\\ \omega^2-V_{RW}&-T_{RW}&0&0\\ -T_{S}&\omega^2-V_{S}&0&0 \end{pmatrix}\,.
\label{Vmatrix}
\end{equation}
The system~\eqref{system} can be solved by standard methods (see e.g. Ref.~\cite{green_funct}). For this purpose, define
the $4\times4$ matrix $X$ whose $n$th column contains the $n$th solution of the homogeneous system
$d\mathbf{x}/dr_*+V\mathbf{x}=0$, i.e. $X_{ij}=x_i^{(j)}$, where the $j$ index denotes a solution of the homogeneous
system and $i$ is the vector index. It can be shown that also the matrix $X$ constructed in such a way is a solution of
the associated homogeneous system, in the sense that
\begin{equation}
  \frac{dX}{d r_*}+VX=0 \label{homosyst}\,.
\end{equation}
In order to solve~\eqref{system}, we impose the ansatz $\mathbf{\Psi}=X\mathbf{\Xi}$,
where $\mathbf{\Xi}$ is a vector to be determined. Substituting the equation above into the inhomogeneous system and
using Eq.~\eqref{homosyst} we find
\begin{equation}
 \frac{d\mathbf{\Xi}}{d r_*}=X^{-1}\mathbf{S}\,,
\end{equation}
and the solution to \eqref{system} reads
\begin{equation}
 \mathbf{\Psi}=X\int dr_* X^{-1} \mathbf{S}\,.\label{particular_sol}
\end{equation}
%

The matrix $X$, contains four independent solutions of the homogeneous system, supplied by suitable boundary
conditions. We impose
\begin{equation}
\left(\begin{array}{c}
  Q\\
  \Theta
\end{array} \right)\to\left(\begin{array}{c}
  A_\pm\\
  B_\pm
\end{array} \right)e^{\pm i \omega r_*}\,,\quad r_*\to\pm\infty\,.\label{BCaxial}
\end{equation}

\noindent As explained in Ref.~\cite{Molina:2010fb}, two linear independent solutions can be constructed by choosing:
(i) $A_\pm=1$ and an arbitrary $B_\pm=B_\pm^{(0)}$ and (ii) $B_\pm=1$ and an arbitrary $A_\pm=A_\pm^{(0)}$, provided
$A_\pm^{(0)}B_\pm^{(0)}\neq1$. This procedure can be applied twice: first we construct two solutions,
$\left\{\mathbf{x}_-^{(1)},\mathbf{x}_-^{(2)}\right\}$, imposing boundary conditions at the horizon and integrating
outward, and secondly we construct two further independent solutions,
$\left\{\mathbf{x}_+^{(1)},\mathbf{x}_+^{(2)}\right\}$, by imposing boundary conditions at infinity and integrating
backward.

Finally, from Eq.~\eqref{particular_sol}, we can write the solutions for the gravitational and scalar waveform, which
satisfy the correct boundary conditions, as follows
\begin{eqnarray}
 Q(r)&=&\sum_{i=1}^2\left(Q_+^{(i)}(r)I_-^{(i)}(r)+Q_-^{(i)}(r)I_+^{(i)}(r)\right)\,,\label{solQ}\\
 \Theta(r)&=&\sum_{i=1}^2\left(\Theta_+^{(i)}(r)I_-^{(i)}(r)+\Theta_-^{(i)}(r)I_+^{(i)}(r)\right)\,,\label{solTheta}
\end{eqnarray}
where
\begin{equation}
 I_\pm^{(i)}=\int_{\pm\infty}^r dr_* \left(C_\pm^{(i)}S_{RW}+D_\pm^{(i)}S_{S}\right)\,,\quad (i=1,2)\nn
\end{equation}
and the functions $C_\pm^{(i)}$ and $D_\pm^{(i)}$  depend on the solutions of the homogeneous system,
$\left\{\mathbf{x}_-^{(1)},\mathbf{x}_-^{(2)}\right\}$ and $\left\{\mathbf{x}_+^{(1)},\mathbf{x}_+^{(2)}\right\}$, and
can be straightforwardly computed from the components of the vector $X^{-1}\mathbf{S}$. For completeness, their expressions are given in Appendix~\ref{app:coeff}.

\subsection{Circular orbits}
So far our approach generically holds for a point-like particle in geodesic motion. However, the above formulae simplify
significantly in the case of circular geodesics at $r=\bar{r}$, as we show in Appendix~\ref{ZRW}. In particular, the
source term for the scalar equation vanishes, $S_s\equiv0$. The remaining source terms can be factorized in order to
extract a Dirac delta contribution $\sim\delta(\omega-m\omega_K)$, where $\omega_K$ is the Keplerian frequency
(\ref{kepl}). In the rest of this Section we write explicitly the indices $\ell,m$ and the dependence on $\omega$.

To compute the Zerilli function we replace Eq.~\eqref{SZomega} in Eq.~\eqref{zerfunc}. Integrating by parts to get rid
of the derivative of the delta function, we get $Z^{\ell m}_\pm(\omega,r)=\bar Z^{\ell
  m}_\pm(r)\delta(\omega-m\omega_K)$ with
\begin{equation} 
\bar Z^{\ell m}_\pm(r)=\frac{Z_\mp(r)}{W_Z}\left[\frac{Z_\pm\hat G^{\ell m}_Z}{f}-\left(\frac{Z_\pm\hat
F^{\ell m}_Z}{f}\right)'\right]_{\bar r}\,,
\nn
\nn
\end{equation} 
for $r\lessgtr\bar r$ respectively; hatted quantities are defined in Appendix~\ref{ZRW}, to which we refer for further
details.
At the boundaries, if we call ${\cal Z}^{\ell m}_\pm(\omega)\equiv
Z^{\ell m}_\pm(\omega,r_*\rightarrow\pm\infty)$, we obtain
\begin{equation}
{\cal Z}^{\ell m}_\pm(\omega)\!=\!\frac{1}{W_Z}\left[\frac{Z_\mp\hat G^{\ell m}_Z}{f}\!-\!\left(\!\frac{Z_\mp\hat
F^{\ell m}_Z}{f}\!\right)'\right]_{\bar r}\!\delta(\omega-m\omega_K)e^{\pm\ii\omega r_*}\,.\nn
\end{equation} 
For later use, we will write this as
\begin{equation}  
{\cal Z}^{\ell m}_\pm(\omega)=\bar {\cal Z}^{\ell m}_\pm\delta(\omega-m\omega_K)e^{\pm\ii\omega r_*}\,,\label{ZbarZ}
\end{equation}  
where $\bar {\cal Z}^{\ell m}_\pm$ is a constant.

A similar procedure can be applied to the axial sector. Using Eq.~\eqref{SRWomega}, the Fourier transform of the
Regge-Wheeler and scalar function, at the boundary $r_*\to \pm\infty$, read
\begin{eqnarray}
 {\cal Q}^{\ell m}_\pm(\omega)&=&\bar {\cal Q}^{\ell  m}_\pm\delta(\omega-m\omega_K)e^{\pm i\omega r_*}\label{QbarQ}\\
 {\cal \varTheta}^{\ell m}_\pm(\omega)&=&\bar {\cal \varTheta}^{\ell  m}_\pm
\delta(\omega-m\omega_K)e^{\pm i\omega r_*}\,,\label{ThbarTh}
\end{eqnarray}
where, from Eqs.~\eqref{solQ} and \eqref{solTheta}, we have
\begin{eqnarray}
 \bar{\cal Q}^{\ell m}_\pm=&&\left\{A_\pm^{(1)}\left[\frac{C_\pm^{(1)}G_{RW}}{f}-\left(\frac{C_\pm^{(1)}
F_{RW}}{f}\right)'\right]_{\bar r}+\right.\nn\\
+&&\left.A_\pm^{(2)}\left[\frac{C_\pm^{(2)}G_{RW}}{f}-\left(\frac{C_\pm^{(2)}F_{RW}}{f}\right)'\right]_{\bar r}\right\}\,,\nn\\
 \bar{\cal \varTheta}^{\ell m}_\pm=&&\left\{B_\pm^{(1)}\left[\frac{C_\pm^{(1)}G_{RW}}{f}-
\left(\frac{C_\pm^{(1)}F_{RW}}{f}\right)'\right]_{\bar r}+\right.\nn\\
+&&\left.B_\pm^{(2)}\left[\frac{C_\pm^{(2)}G_{RW}}{f}-
\left(\frac{C_\pm^{(2)}F_{RW}}{f}\right)'\right]_{\bar r}
\right\}\,.\nn
\end{eqnarray}
As we show in the next section, the energy flux at infinity and at the horizon can be computed in terms of the
quantities ${\cal Z}^{\ell m}_\pm$, ${\cal Q}^{\ell m}_\pm$ and $\varTheta^{\ell m}_\pm$ given in Eqs.~\eqref{ZbarZ},
\eqref{QbarQ} and \eqref{ThbarTh}, respectively.
\section{Results}\label{results}
\subsection{Energy flux}\label{flux}
The flux of gravitational energy can be computed in terms of metric perturbations.  At asymptotically flat, future,
null infinity, the expressions of the effective gravitational wave stress-energy tensor (i.e. the Isaacson tensor) in
DCS gravity and in general relativity coincide \cite{Stein:2010pn}. Therefore, we can use the machinery derived in the
framework of general relativity to determine the emitted gravitational energy flux at infinity in DCS gravity.

On the other hand, the expression of the rate of energy absorbed by the horizon is also formally equivalent to that in
general relativity (see e.g. Ref.~\cite{Poisson:2004cw}). Indeed, the derivation involves the first law of BH
thermodynamics, which relates the change in energy $\dot M$ with the change in the horizon area $\dot A$, and
Raychaudhuri's equation to calculate $\dot A$ in terms of the shear tensor. In DCS gravity, both these steps proceed
exactly as in general relativity, all the dynamical information being eventually encoded in the waveforms.

Hence, the energy fluxes at (null) infinity and at the horizon formally read as in general relativity
\cite{Martel:2005ir}
\begin{eqnarray}
\dot E_{grav}^\pm&\equiv&\left<\frac{dE_{grav}}{dx}\right>\nonumber\\
&=&\frac{1}{64\pi}\frac{(\ell+2)!}{(\ell-2)!}\sum_{\ell m}\left[|{\dot
{\cal Z}_\pm}^{\ell m}(x)|^2+4|{{\cal Q}}_\pm^{\ell m}(x)|^2\right]\,,\nonumber\\
\end{eqnarray}
where the sum is taken over negative and positive $m$ and $x=t\mp r_*$ are the retarded and advanced coordinates, respectively.  The inverse Fourier transform of
(\ref{ZbarZ}), (\ref{QbarQ}) is
\begin{eqnarray}
{\cal Z}^{\ell m}_\pm(t\mp r_*)&=&\int d\omega\bar{{\cal Z}}_\pm\delta(\omega-m\omega_K)e^{-\ii m\omega_K(t\mp r_*)}\nn\\
&=&{\bar
{\cal Z}}^{\ell m}_\pm e^{-\ii m\omega_K(t\mp r_*)}\nonumber\\
{\cal Q}^{\ell m}_\pm(t\mp r_*)&=&\int d\omega\bar {\cal Q}_\pm\delta(\omega-m\omega_K)e^{-\ii m\omega_K(t\mp r_*)}\nn\\
&=&{\bar
{\cal Q}}^{\ell m}_\pm e^{-\ii m\omega_K(t\mp r_*)}\,,\nn
\end{eqnarray}
therefore
\begin{equation}
\dot E_{grav}^\pm=\frac{1}{64\pi}\frac{(\ell+2)!}{(\ell-2)!}\sum_{\ell m}\left[(m\omega_K)^2\left|{\bar
{\cal Z}}^{\ell m}_\pm\right|^2+4\left|{\bar {\cal Q}}^{\ell m}_\pm\right|^2\right]\,.\label{flux_grav}
\end{equation}

On the other hand, the scalar energy flux reads (see, e.g. Ref.~\cite{Martel:2003jj})
\begin{equation}
\dot E_{scal}=-r^2f(r)\int d\Omega\, T_{tr}^\text{scal}\,.
\end{equation} 
From the stress-energy tensor of the scalar field,
$T_{ab}^\text{scal}=\beta(\nabla_{(a}\vartheta^*\nabla_{b)}\vartheta-1/2g_{ab}\nabla_c\vartheta\nabla^c\vartheta^*)$. Inserting
Eq.~\eqref{scalar_decomp} and using the asymptotic behavior at infinity (\ref{ThbarTh}), the energy flux reads
\begin{equation}
\dot E_{scal}^\pm\equiv\left<\frac{dE_{scal}}{dx}\right>=\sum_{\ell m}(m\omega_K)^2\beta\left|\bar\varTheta^{\ell m}_\pm\right|^2\,.\label{flux_scalar}
\end{equation}
\begin{figure*}[htb]
\begin{center}
\begin{tabular}{cc}
\epsfig{file=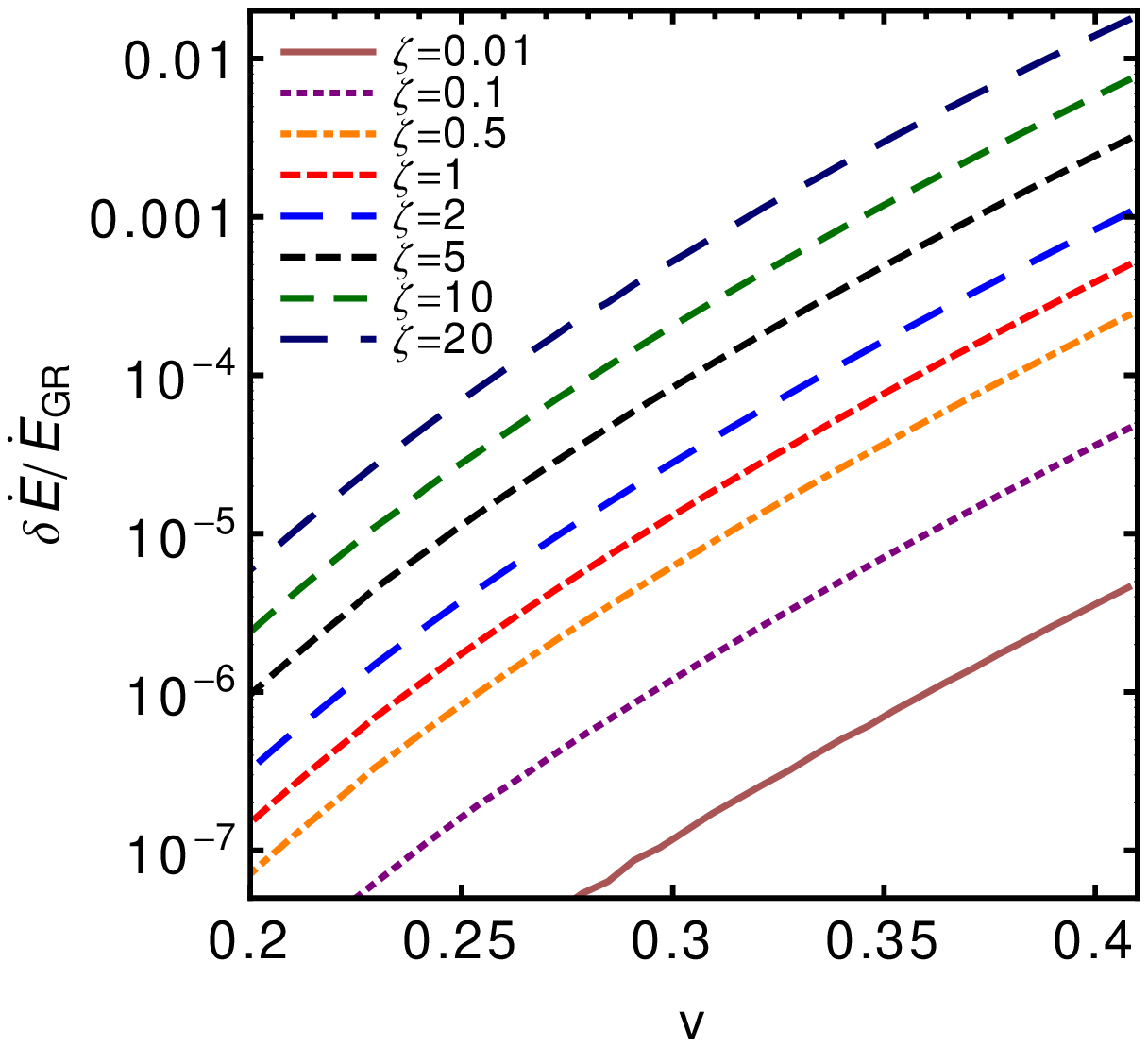,height=6.5cm,angle=0}&
\epsfig{file=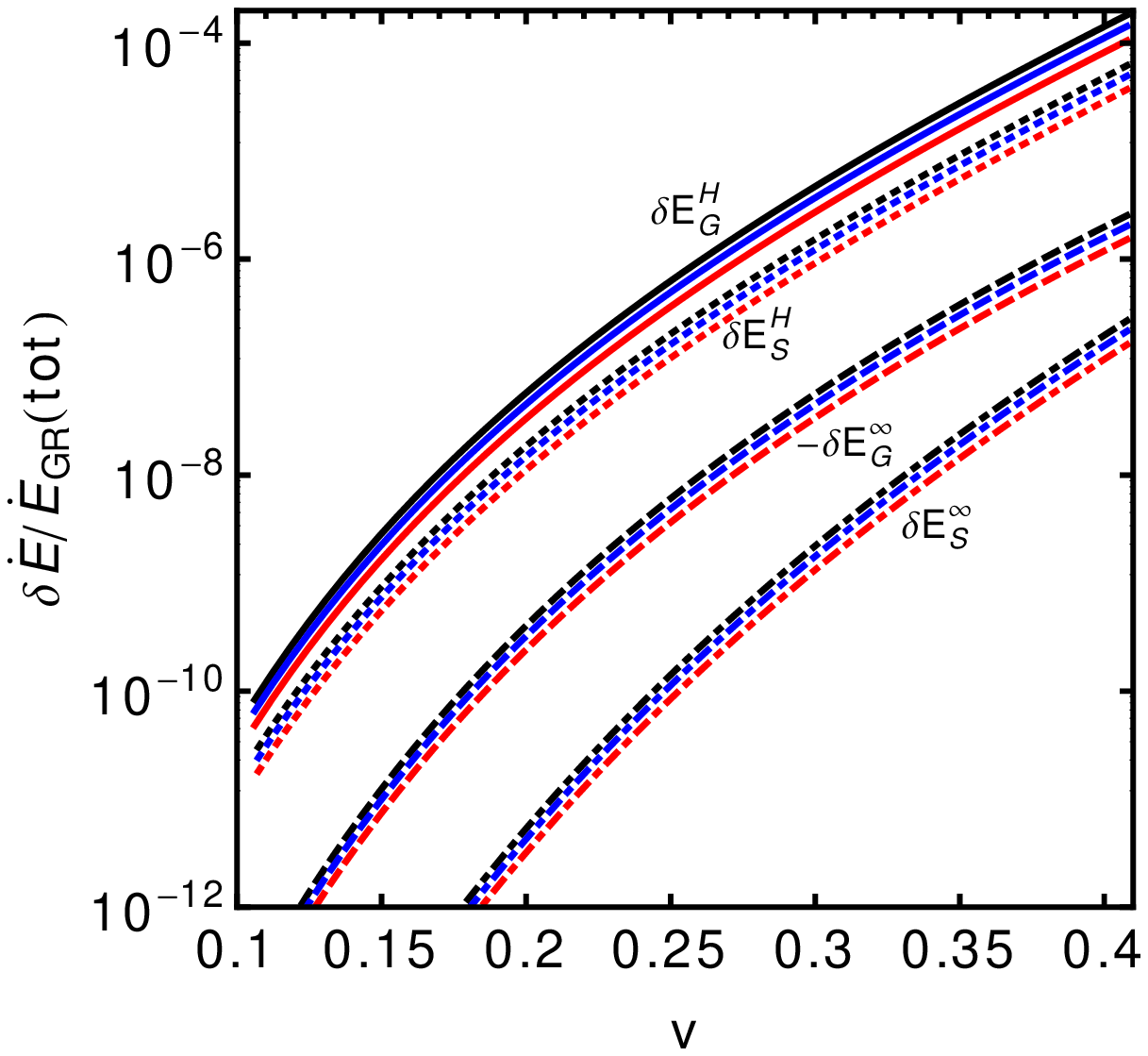,height=6.5cm,angle=0}
\end{tabular}
\caption{(Color online) Left: Relative difference between the power emitted in gravitational waves in general relativity and DCS gravity,
cf. Eq.~\eqref{relativeP}, for different values of $\zeta$. The sum is truncated at $\ell=3$.  The contribution in DCS
gravity includes both the gravitational and the scalar fluxes, 
$\dot{E}_\text{DCS}=\dot{E}_\text{grav}+\dot{E}_\text{scal}$ 
at infinity and at the horizon.  Right: 
Different contributions to the relative difference for different values of $\zeta=0.5,0.4,0.3$, corresponding to
different colors. The main contributions arise from the fluxes at the horizon and they are \emph{positive}, whereas the difference in the gravitational flux at infinity is negative.  
\label{fig:deltaP}}
\end{center}
\end{figure*}

Finally, since the orbital frequency is related to the orbital velocity $v$ and to the semi-latus rectum (which for
circular orbits is simply $p=\bar r/M$) by the relations
\begin{equation}
v=\left( M\omega_K\right)^{1/3}=p^{-1/2}\,,
\end{equation}
the energy flux $\dot E$ can also be considered either as a function of $v$ or $p$. The condition for the existence of
stable circular orbits, $\bar{r}>r_\text{ISCO}=6M$, constrains the values of $v$ and $p$ to $p>6$ and
$v<6^{-1/2}\sim0.408$.

The method described above has been implemented in \textsc{Mathematica}. In our numerical approach, we have considered a series
expansion at the horizon and at infinity up to order eight for the boundary conditions~\eqref{BCpolar}
and~\eqref{BCaxial}. Our results are summarized
in Fig.~\ref{fig:deltaP}. When $\zeta=0$, our results agree with those in general
relativity~\cite{Cutler:1993vq,Poisson:1995vs} within one part in $10^6$ or better. Furthermore, in the small $\zeta$
limit, we develop an independent method (discussed in Appendix~\ref{app:green}) whose results are in perfect agreement
with the ones discussed here.

As expected, the CS corrections are more effective when $p\sim6$, i.e. close to the innermost stable circular orbit
(ISCO), where circular orbits probe the strong curvature region around the massive BH.  Far away from the source the CS
contributions are negligible. This is clear from the left panel of Fig.~\ref{fig:deltaP}, where we show the relative
difference in the emitted power
\begin{equation}
\frac{\delta \dot{E}}{\dot{E}_\text{GR}}\equiv \frac{\dot{E}_\text{DCS}-\dot{E}_\text{GR}}{\dot{E}_\text{GR}}\,,\label{relativeP}
\end{equation}
where $\dot{E}_\text{DCS}=\dot{E}_{grav}^H+\dot{E}_{grav}^\infty+\dot{E}_{scal}^H+\dot{E}_{scal}^\infty$, i.e. it is the sum of the contributions
coming from the gravitational and scalar fluxes, both at the infinity and the horizon, and 
$\dot{E}_\text{GR}=\dot{E}_\text{DCS}(\zeta=0)$, i.e. the energy flux in general relativity. Clearly, the scalar contribution
to $\dot{E}_\text{GR}$ is vanishing. The relative difference is positive, i.e. the total power
emitted in DCS gravity is \emph{larger} than in general relativity. This is consistent with the fact that in this theory
there is an extra scalar degree of freedom, which introduces further energy dissipation channels. Although the
difference in the \emph{total} flux is positive, we find that for some subdominant ($\ell\geq3$) mode, the energy flux
may be smaller than the corresponding flux in general relativity. This shows that a conversion of scalar into
gravitational energy is possible, due to the CS coupling.

Furthermore, even if the axial flux can be as large as twice the axial flux in general relativity (for example when
$p\sim6$ and for $\zeta\sim10$), the correction to the total energy flux is significantly smaller. Indeed, the leading
contribution to the energy flux arises from $\ell=m$ modes, which have polar parity because selection rules imply that
for even values of $\ell+m$ only polar perturbations are sourced (see Appendix~\ref{ZRW} for details). Therefore, since
DCS corrections only affect the axial sector of Schwarzschild perturbations, their contribution is subleading with
respect to that coming from the polar perturbations. In the most favorable case ($p\sim6$ and $\zeta\sim10$) the total
energy flux (summing over polar and axial contributions up to $\ell=5$ and $-l\leq m\leq l$) only differs from the
general relativity value by a few percent. Typically, the deviation is smaller, as shown in the left panel of
Fig.~\ref{fig:deltaP}.

In the right panel of Fig.~\ref{fig:deltaP} we show the four contributions to the total emitted power.  Remarkably, the
main contributions arise from the gravitational and scalar flux \emph{at the horizon}, which are positive and sensibly
larger than the contributions at infinity; see Appendix \ref{app:green} for a discussion on this behaviour.

Note that the correction to the gravitational flux at infinity is negative but, since this is a subleading contribution,
the correction to the total energy flux is nevertheless positive.

Finally, from the results shown in Fig.~\ref{fig:deltaP} we can extract the following dependence in the small $v$ limit
\begin{eqnarray}
 \frac{\delta\dot{E}_{grav}^H}{\dot E_\text{GR}^\text{tot}}&\sim& v^{10}  \,,\qquad \frac{\delta\dot{E}_{scal}^H}{\dot E_\text{GR}^\text{tot}}\sim v^{10}\,,\\
 \frac{\delta \dot{E}_{grav}^\infty}{\dot E_\text{GR}^\text{tot}}&\sim& v^{12}\,,\qquad 
 \frac{\delta \dot{E}_{scal}^\infty}{\dot E_\text{GR}^\text{tot}}\sim v^{14}\,.\label{pnresult}
\end{eqnarray}
Notice that the high power in $v$ introduces large errors in the fits above and these results should be understood as
\emph{lower limits} for the post-Newtonian (PN) orders of the CS effects. The scalar flux at infinity is consistent with
analytical predictions at PN level~\cite{Yunes:preliminary}.

\subsection{Gravitational-wave signal}\label{dcycles}
Although the difference between the energy emission by EMRIs in general relativity and their emission in DCS gravity is
small, stellar-mass objects can orbit supermassive BHs for $\sim 10^5$ cycles (many of which can occur near the ISCO)
while in the sensitivity window of LISA, before reaching the ISCO and eventually plunge. Hence, the small deviations
accumulate and they can result in sensible modifications when one looks at the entire inspiral$+$merger process. Let us
give a rough estimate of this effect, and of its observability by space based detectors like LISA.

A useful quantity to consider is the number of gravitational-wave cycles accumulated within a certain frequency band
(see for instance Ref.~\cite{Berti:2004bd}).  This quantity is defined as\footnote{In this section $f$
 indicates the gravitational wave frequency, whereas, in the previous sections, it denoted the $\{t,t\}$ component of the
 metric~\eqref{schwmet}. To avoid confusion while keeping the standard notation, in Eq.~\eqref{frc} this component is
 denoted by $g_{00}(r)$.}
\begin{equation}
{\cal N}=\int_{f_i}^{f_f}\frac{f}{\dot f}df\,,\label{cycles}
\end{equation}
with
\begin{eqnarray} 
f_i&=&\max\left(f_\text{low},f_\text{1yr}\right)\,,\\ 
f_f&=&\min\left(f_\text{ISCO},f_\text{up}\right)\,.
\end{eqnarray}
\begin{table*}[th!]
\caption{\label{tab:cycles} Corrections to the number of gravitational wave cycles accumulated within the frequency
band $f\in[f_i,f_f]$ for some typical two-body systems. Data correspond to $T_\text{obs}=1 $yr and are roughly fitted 
by Eq.~\eqref{fit}.}
\begin{tabular}{c|cc|cc|cc|cc|cc}
M &\multicolumn{2}{c}{$(1.4+10^4)M_\odot$} & \multicolumn{2}{c}{$(1.4+7\times10^5)M_\odot$} & 
\multicolumn{2}{c}{$(1.4+4\times10^6)M_\odot$} & \multicolumn{2}{c}{$(10+4\times10^6)M_\odot$} & 
\multicolumn{2}{c}{$(10+10^7)M_\odot$}\\
$f_i$(Hz)&\multicolumn{2}{c}{0.0205} & \multicolumn{2}{c}{0.0041} & \multicolumn{2}{c}{0.0010} & 
\multicolumn{2}{c}{0.00088} & \multicolumn{2}{c}{0.00040}\\
$f_f$(Hz)&\multicolumn{2}{c}{0.4396} & \multicolumn{2}{c}{0.0063} & \multicolumn{2}{c}{0.0011} & 
\multicolumn{2}{c}{0.00110} & \multicolumn{2}{c}{0.00044}\\
\hline \hline
$\zeta$ & $-\delta{\cal N}$ &  $-\delta {\cal N}/{\cal N}$ & $-\delta{\cal N}$ &  
$-\delta {\cal N}/{\cal N}$ & $-\delta{\cal N}$ &  $-\delta {\cal N}/{\cal N}$ 
& $-\delta{\cal N}$ &  $-\delta {\cal N}/{\cal N}$ & $-\delta{\cal N}$ &  $-\delta {\cal N}/{\cal N}$  \\
\hline 
20 	&	38.2	&	$3.78\times 10^{-5}$	&	900	&	$6.12\times 10^{-3}$	&	455	&	$1.40\times 10^{-2}$ &	295	&	$9.94\times 10^{-3}$ &	179	&	$1.38\times 10^{-2}$	\\
10 	&	15.2	&	$1.51\times 10^{-5}$	&	365	&	$2.48\times 10^{-3}$	&	188	&	$5.78\times 10^{-3}$ &	121	&	$4.07\times 10^{-3}$ &	73.9	&	$5.69\times 10^{-3}$	\\
5 	&	6.28	&	$6.21\times 10^{-6}$	&	152	&	$1.04\times 10^{-3}$	&	79.5	&	$2.44\times 10^{-3}$ &	50.8	&	$1.71\times 10^{-3}$ &	31.2	&	$2.40\times 10^{-3}$	\\
2 	&	2.12	&	$2.09\times 10^{-6}$	&	51.7	&	$3.51\times 10^{-4}$	&	27.2	&	$8.34\times 10^{-4}$ &	17.3	&	$5.82\times 10^{-4}$ &	10.7	&	$8.20\times 10^{-4}$	\\
1 	&	0.98	&	$9.70\times 10^{-7}$	&	24.0	&	$1.64\times 10^{-4}$	&	12.7	&	$3.89\times 10^{-4}$ &	8.05	&	$2.71\times 10^{-4}$ &	4.97	&	$3.82\times 10^{-4}$	\\
0.5 	&	0.47	&	$4.65\times 10^{-7}$	&	11.5	&	$7.85\times 10^{-5}$	&	6.09	&	$1.87\times 10^{-4}$ &	3.87	&	$1.30\times 10^{-4}$ &	2.39	&	$1.84\times 10^{-4}$	\\
0.4 	&	0.37	&	$3.69\times 10^{-7}$	&	9.15	&	$6.23\times 10^{-5}$	&	4.83	&	$1.48\times 10^{-4}$ &	3.07	&	$1.00\times 10^{-4}$ &	1.89	&	$1.45\times 10^{-4}$	\\
0.3 	&	0.28	&	$2.74\times 10^{-7}$	&	6.81	&	$4.63\times 10^{-5}$	&	3.59	&	$1.10\times 10^{-4}$ &	2.28	&	$7.68\times 10^{-5}$ &	1.40	&	$1.08\times 10^{-4}$	\\
0.2 	&	0.18	&	$1.81\times 10^{-7}$	&	4.50	&	$3.06\times 10^{-5}$	&	2.37	&	$7.29\times 10^{-4}$ &	1.51	&	$5.08\times 10^{-5}$ &	0.93	&	$7.17\times 10^{-5}$	\\
0.1 	&	0.09	&	$8.97\times 10^{-8}$	&	2.23	&	$1.52\times 10^{-5}$	&	1.18	&	$3.61\times 10^{-5}$ &	0.75	&	$2.52\times 10^{-5}$ &	0.46	&	$3.55\times 10^{-5}$	\\
0.01 	&	0.009	&	$8.95\times 10^{-9}$	&	0.221	&	$1.51\times 10^{-6}$	&	0.12	&	$3.61\times 10^{-6}$ &	0.074	&	$2.50\times 10^{-6}$ &	0.046	&	$3.53\times 10^{-6}$	\\
\hline
\hline
\end{tabular}
\end{table*}
%
\begin{figure*}[htb]
\begin{center}
\begin{tabular}{cc}
\epsfig{file=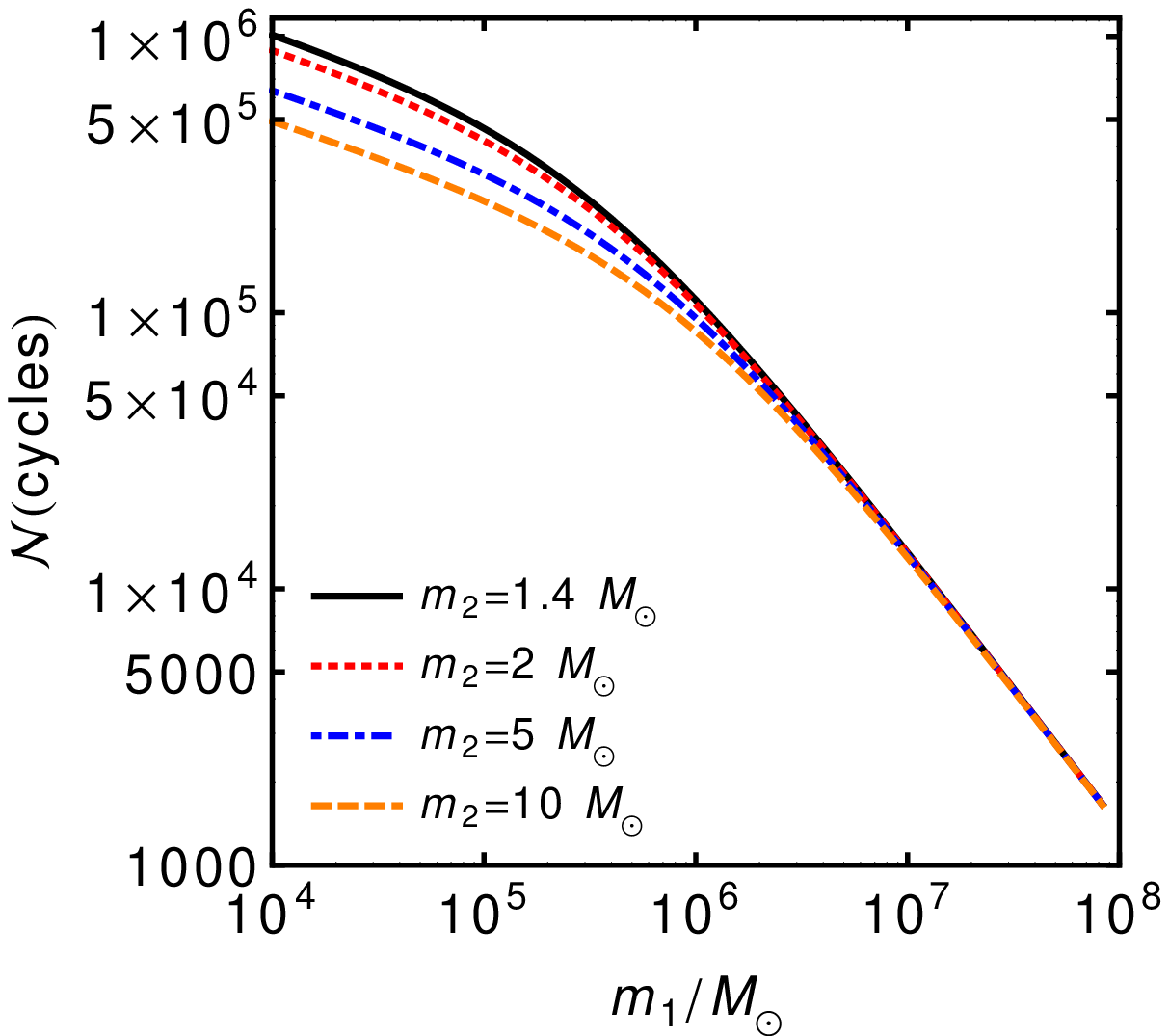,height=6.5cm,angle=0}&
\epsfig{file=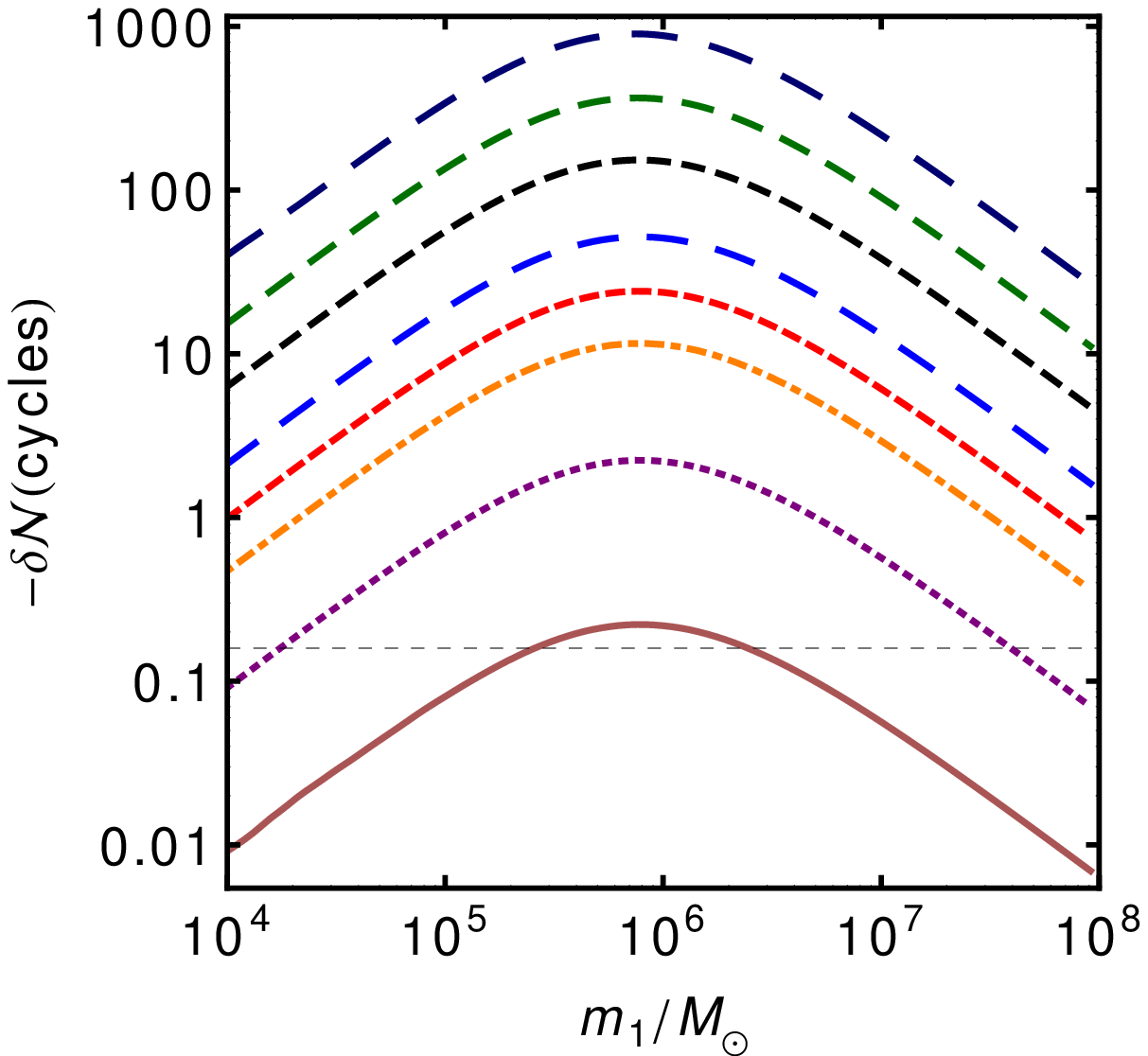,height=6.5cm,angle=0}
\end{tabular}
\caption{(Color online) Left: Total number of cycles ${\cal N}$ as a function of the central mass $m_1$, 
for several values of $m_2$, $T_\text{obs}=1$yr and $\zeta=0.5$.
Right: Corrections to the number of gravitational wave cycles accumulated during the inspiral of a small object ($m_2=1.4\,M_\odot$) 
around a supermassive BH of mass $m_1$ in 1 year observation time before coalescence. We show $\delta{\cal N}$ 
as a function of $m_1$ for different values of $\zeta$. Same legend as in Fig.~\ref{fig:deltaP}
\label{fig:deltaN}}
\end{center}
\end{figure*}
\begin{figure}[htb]
\begin{center}
\begin{tabular}{c}
\epsfig{file=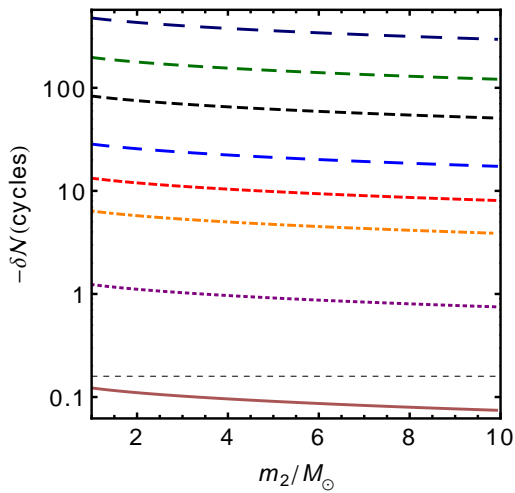,height=6.5cm,angle=0}
\end{tabular}
\caption{(Color online) 
Correction to the number of gravitational wave cycles accumulated during the inspiral of a small object around a supermassive 
BH of mass $m_1=4\times 10^6M_\odot$ in 1 year observation time before coalescence, as a function of the mass $m_2$ 
of the small object and different values of $\zeta$ (same legend as in Fig.~\ref{fig:deltaP}). 
\label{fig:deltaN_m2}}
\end{center}
\end{figure}
\begin{figure*}[htb]
\begin{center}
\begin{tabular}{cc}
\epsfig{file=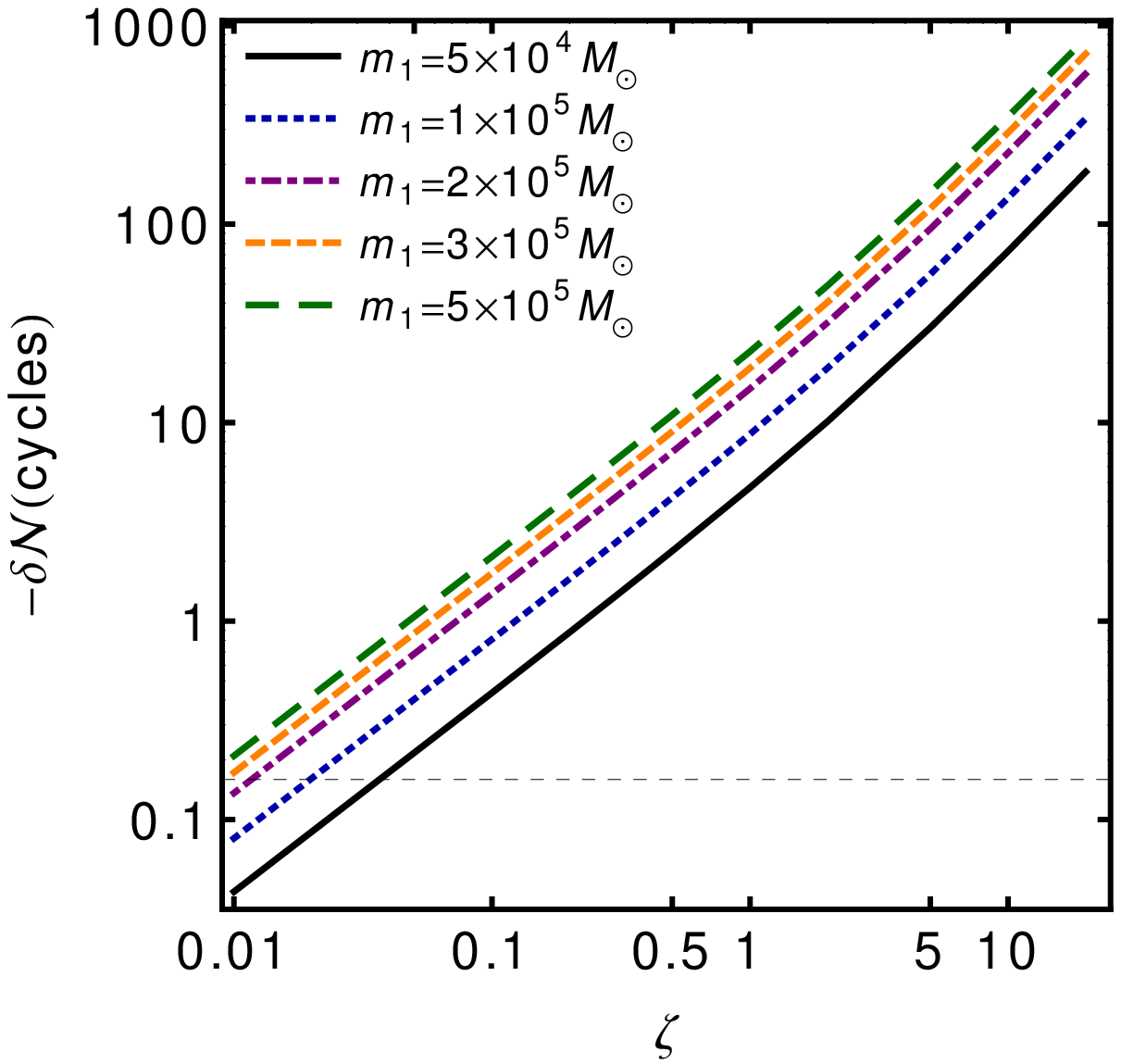,height=6.5cm,angle=0}&
\epsfig{file=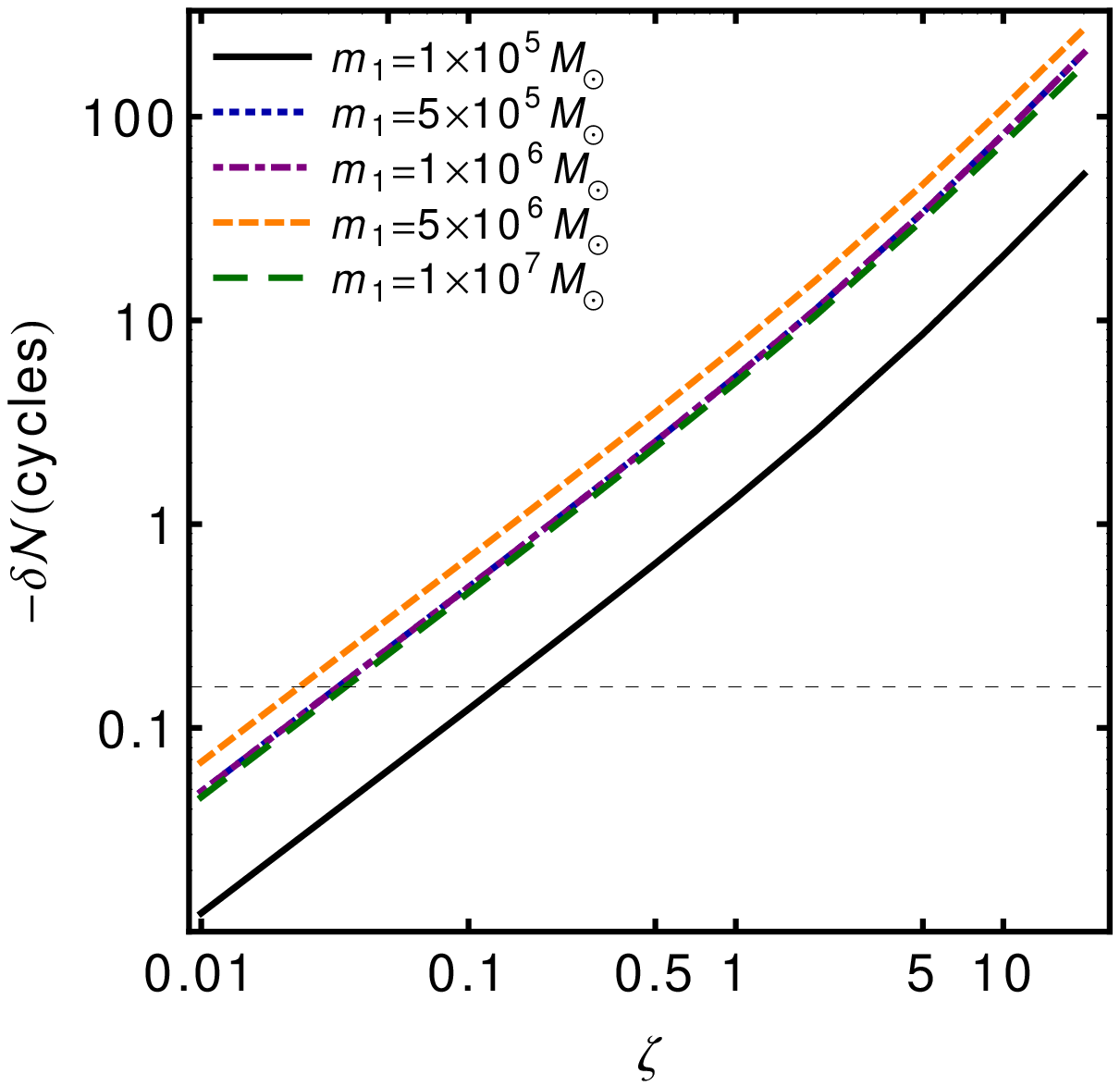,height=6.5cm,angle=0}
\end{tabular}
\caption{(Color online) Correction to the number of gravitational wave cycles accumulated during the inspiral of a small object around
a supermassive BH of mass $m_1$ in 1 year observation time before coalescence. Left: $\delta{\cal N}$ as a 
function of $\zeta$ for some values of $m_1$ and $m_2=1.4M_\odot$. Right: same with $m_2=10 M_\odot$. The corrections 
are linear in $\zeta$.\label{fig:deltaNVSzeta}}
\end{center}
\end{figure*}

Here $f_\text{low}=10^{-5}$ Hz and $f_\text{up}=1\text{ Hz}$ are two typical (lower and upper, respectively) cutoffs of 
the LISA noise curve, $f_\text{ISCO}=(6^{3/2}\pi M)^{-1}$ is the frequency at the ISCO.  Finally, $f_\text{1yr}$ is the 
frequency of the gravitational radiation emitted one year before the coalescence (coalescence is for simplicity defined 
to occur when the particle reaches the ISCO).  This choice is due to the fact that, despite the timespan of the LISA 
mission will likely be $\sim1\!-\!5$ years, we make the most conservative assumption, considering one year of 
observation time.  The frequency $f_\text{1yr}$ is obtained by solving
\begin{equation} 
T_\text{obs}=\int_{f_\text{1yr}}^{f_\text{ISCO}}\frac{df}{\dot{f}}=1 \text{ year}\,.\label{time}
\end{equation}
To solve Eqs.~(\ref{cycles}) and (\ref{time}), one has to know how the particle is inspiralling into the massive BH.  We
use a simple prescription, the so-called adiabatic approximation, in which the back-reaction is obtained by the energy
flux, that is computed assuming that the particle moves on a geodesic orbit. Later in this Section we shall discuss this
approach in more detail, assessing its accuracy; here we only remark that the error due to this approximation is much
smaller than the other uncertainties in our problem (specifically the observation time, but also fundamental issues such
as the magnitude of the coupling constants).

In order to compute $f_\text{1yr}$ using Eq.~(\ref{time}), one may consider a PN formula for $\dot f$. This choice has
been widely adopted in previous works~\cite{Berti:2004bd,Blanchet:1995ez,Ohashi:1996uz}. For EMRIs, the velocity during
the latest stages before coalescence can be as large as a fraction of the speed of light and the PN approach may
introduce sensible errors, as discussed in Appendix \ref{app:comparison}.  Therefore, we adopt a different prescription,
as explained below. Our approach is well suited for the large-velocity and strong-field regime and thus it is expected
to be more accurate for EMRIs. In Appendix \ref{app:comparison} we compare it with other, less accurate, approaches.

We consider geodesic motion of a two-body system with masses $m_1$ and $m_2$. The motion can be effectively reduced to
that of a particle with reduced mass $\mu=m_1m_2/M=\eta M$ orbiting a central object with mass $M=m_1+m_2$, consistently with the notation used in the previous sections.
The frequency of gravitational waves emitted by the particle on a circular geodesic at $r=\bar r$ reads
\begin{equation}
f=\frac{\omega_K}{\pi}=\frac{1}{\pi}\sqrt{\frac{g_{00}'({\bar r})}{2{\bar r}}}=\frac{1}{\pi}\sqrt{\frac{M}{{\bar r}^3}}\,,\label{frc}
\end{equation}
from which we get $\dot f=-\frac{3}{2}f\frac{\dot {\bar r}}{{\bar r}}$. For a particle in a circular orbit we have
\begin{equation}
E_{orb}=\frac{{\bar r}-2M}{\sqrt{{\bar r}({\bar r}-3M)}}\mu\,.
\end{equation}
Putting all together, we find
\begin{equation}
\dot f=-\frac{3}{2}\frac{f}{{\bar r}}\frac{d{\bar r}}{dE_{orb}}\dot E_{orb}\,.\label{fdotGR}
\end{equation}
Finally, using Eq.~\eqref{frc} we get
\begin{eqnarray}
\dot f=\frac{3}{\pi^{2/3}\mu M^{5/6}}\frac{\left[M^{1/3}f^{-2/3}-3M\pi^{2/3}\right]^{3/2}}{\left[M^{1/3}f^{-2/3}-
6M\pi^{2/3}\right]}f^{2/3}\dot E_\text{DCS}\nonumber\\
\label{dotfFIN}
\end{eqnarray}
where $\dot E_\text{DCS}=\dot E_\text{grav}+\dot E_\text{scal}=-\dot E_{orb}$
is the total energy flux radiated away, using 
the standard flux balance equation
\begin{equation}
\dot E_{orb}+\dot E_{grav}+\dot E_{scal}=0\,.\label{balance}
\end{equation}
In the balance equation above, both the contributions at the horizon and at infinity must be taken into account.

We remark that we are modeling the EMRI orbit in the adiabatic approximation: the particle is in nearly geodesic motion,
allowing to compute, at each time, the emitted energy flux $\dot E_{DCS}$ assuming a geodesic orbit; furthermore, in this
approach one assumes that the back-reaction is described by the flux balance (\ref{balance}).  This approximation
neglects the so-called ``conservative part of the self-force'', which is a higher order effect contributing marginally
to the gravitational signal (see Sec.~2 of \cite{2011GWN.....5....3T}).  Indeed, as shown in \cite{Huerta:2008gb}, for
spinning BHs the conservative contribution can account for at most few cycles of the entire process, while for
non-spinning BHs it contributes less than one radiant to a one year evolution~\cite{Yunes:2009ef}.  Therefore, it may be
necessary to take it into account in the data analysis of the process, but this effect can be neglected in assessing the
relevance of DCS corrections for the EMRI signal.

Using the energy fluxes computed in the previous Section, we can therefore obtain the frequency at any instant prior to
merger and the number of cycles ${\cal N}$ left to merger. The total number of cycles is shown in the left panel of Fig.~\ref{fig:deltaN}.

We can also compute the correction to the total number of cycles due to the CS coupling as follows.
From the relation (\ref{dotfFIN}),  
we obtain $\delta \dot f/\dot f=\delta \dot E/\dot E_{GR}$, with $\delta X=X_\text{DCS}-X_\text{GR}$. 
Then, at first order,
\begin{equation}
{\cal N}=\int_{f_i}^{f_f}\frac{f}{\dot f}\left[1-\frac{\delta\dot E}{\dot E_{GR}}\right]df\,,
\end{equation}
and the correction in the number of cycles reads
\begin{equation}
\frac{\delta {\cal N}}{{\cal N}}=-\frac{\int_{f_i}^{f_f}\frac{f}{\dot f}
\frac{\delta\dot E}{\dot E_{GR}}df}{\int_{f_i}^{f_f}\frac{f}{\dot f}df}\,.\label{correctionN}
\end{equation}
The difference in the number of cycles is shown in Table~\ref{tab:cycles} for different
values of the CS coupling $\zeta$ and for a typical set of parameters.

The corrections depend on $m_1$, $m_2$ and on the CS coupling $\zeta$.  This is shown in the right panel of
Fig.~\ref{fig:deltaN} and in Fig.~\ref{fig:deltaN_m2}, where we indicate the fiducial threshold $\delta{\cal
  N}=(2\pi)^{-1}$ cycles (i.e. $\delta\Phi=1$ rad) with a horizontal line. Corrections to general relativity
are generally considered significant if they exceed one radiant over the observation time \cite{Cutler:1992tc}.

In the right panel of Fig.~\ref{fig:deltaN_m2} we show the dependence on $m_2$ and on $\zeta$ for a central supermassive
object with $m_1=4\times 10^6 M_\odot$ (i.e., the mass of the supermassive BH at the center of the Milky
Way~\cite{Ghez:2008ms}). Note that the dependence on $m_2$ appears to be very mild.

Overall, our results are well described by
\begin{equation}
\delta{\cal N}\sim-26\zeta\sqrt{\frac{M_\odot}{m_2}}\exp\left\{-1.2\log_{10}^2\left[\frac{m_1}{m_\text{max}}\right]\right\}\,,\label{fit}
\end{equation}
where
\begin{equation}
m_\text{max}=6.6\times 10^5 M_\odot\sqrt{\frac{m_2}{M_\odot}}\,,\label{mmax}
\end{equation}
is the location of the maximum in the left panel of Fig.~\ref{fig:deltaN} and it does not depend on the CS coupling.
The fit above has been inspired by the curves in the left panel of Fig.~\ref{fig:deltaN}. In a semi-logarithmic scale,
these curves are approximately Gaussian, $y=y_0 \exp\left[{a_0(x-x_m)^2}\right]$, where $y_0$, $a_0$ and $x_m$ are the
fit parameters and the expression~\eqref{fit} is simply written in the coordinate $x=\log_{10}(m_1)$.  We estimate an
error on the fit smaller than a few percent when $m_1\in[10^5,10^7]M_\odot$, $m_2\in[1,10]M_\odot$ and $\zeta\le1$.
As shown in Fig.~\ref{fig:deltaNVSzeta}, for larger values of the CS coupling, $|\delta {\cal N}|$ grows faster than linearly as a function of $\zeta$, and Eq.~\eqref{fit} would acquire higher order in $\zeta^2$ contributions. For example when $\zeta\sim20$, the fit~\eqref{fit} is accurate within $50\%$.

The presence of a maximum in $\delta{\cal N}(m_1)$ (Fig.~\ref{fig:deltaN}, right panel) can be understood as
follows: on one hand, the DCS correction for a given value of $\zeta=16\pi\alpha^2/(\beta M^4)$ becomes more significant
as $M=m_1+m_2$ increases, as it appears from Eqs.~(\ref{eqRW})-(\ref{potentialZer}); on the other hand, for
large values of $M$ the total number of cycles decreases (see Fig.~\ref{fig:deltaN}, left panel), and thus $\delta {\cal
  N}$ decreases, too.

\section{Conclusions}\label{concl}
We have studied the gravitational wave emission by a small object on a quasi-circular geodesic around a static,
spherically symmetric, massive BH, in the context of Dynamical Chern-Simons gravity. This process can describe for
instance the inspiralling of a neutron star or stellar-mass BH into a supermassive BH, and is thought to occur
frequently in the universe. In fact, EMRIs are one of the main preferred sources of gravitational waves for space-based
detector LISA. We have shown that, because the stellar-size object spends many cycles in the bandwidth of LISA, the
small effect of coupling to the DCS term ``piles up'' giving rise to measurable effects, in particular a decrease in the
number of cycles over a fixed frequency bandwidth.

Extensions of this work are necessary before useful constraints to the theory can be obtained.  In particular, the space
of Schwarzschild BHs is a set of measure zero in the space of solutions: real BHs are most likely to be rotating. An
extension of the present formalism and results to spinning BHs (for instance to the general class found in
Ref.~\cite{Vigeland:2011ji}) is highly desirable, but might have to wait for more powerful techniques or full-blown
numerical simulations. Since the curvature invariants are higher close to rapidly spinning BHs and since the ISCO
gets closer to the horizon, rapidly spinning BHs are a potentially very interesting tool to test DCS gravity.
Furthermore, we have focused on circular orbits, but eccentricity may play an important role. The formalism we developed
already allows for studies of eccentric EMRIs in DCS gravity.
For eccentric orbits around rotating BHs much larger corrections due to the CS coupling are expected. Therefore, our
results should be seen as a lower limit of DCS corrections to the EMRI gravitational-wave signal.

Finally, the results described in this paper are valid for any value of $\zeta$. However, astrophysical observations
already constrain the CS coupling~\cite{Yunes:2009hc}
\begin{equation}
\xi=\frac{16\pi\alpha^2}{\beta}\lesssim 10^{16}\text{km}^4
\,. \label{bound}
\end{equation}
Notice that $\alpha$ and $\beta$ are the \emph{physical} parameters entering the action~\eqref{action}, so that the constraints on $\zeta$
(for instance those possibly arising from a comparison between future observational data from LISA
with Table~\ref{tab:cycles} or Figs.~\ref{fig:deltaN}-\ref{fig:deltaNVSzeta}) should always be
converted into constraints on $\xi$, depending on the lengthscale of the system under consideration.

In this case, from the definition~\eqref{zetadef} and the
constraint~\eqref{bound} we obtain that, for objects with mass $M\lesssim 10^4 M_\odot$, the CS coupling
$\zeta$ may be larger than unit and our results show that in this case the CS coupling introduces large corrections,
which are potentially detectable and not already ruled out by previous constraints. However, the point-like
approximation used in this paper is justified for EMRIs ($m_1/m_2\sim 10^4$ or larger) whereas, for smaller values of
the mass ratio the structure of the smaller object, and the details of its backreaction, should be taken into account.

\begin{acknowledgments} 
  It is our pleasure to thank Nico Yunes and Emanuele Berti for enlightening conversations and Eric Poisson for useful
  correspondence.  We also thank Kent Yagi, Leo Stein and Nico Yunes for sharing with us some of their preliminary
  results~\cite{Yunes:preliminary}.  This work was supported by the {\it DyBHo--256667} ERC Starting Grant and by FCT -
  Portugal through PTDC projects FIS/098025/2008, FIS/098032/2008, CTE-AST/098034/2008 and CERN/FP/116341/2010.
\end{acknowledgments}
\appendix
\section{Derivation of the perturbation equations}\label{ZRW}
In this appendix we describe the derivation of the perturbation equations~\eqref{eqRW},
\eqref{eqScalar} and \eqref{eqZer} with source terms $S_\text{RW}$, $S_S$ and $S_Z$ respectively.

Let us start with some definitions. We denote the Regge-Wheeler (RW) function by $Q$ and the Zerilli function by $Z$.
$Y^{\ell m}$ are the usual scalar spherical harmonics, in term of which, following
Refs.~\cite{Martel:2003jj,Martel:2005ir} we define vector and tensor spherical harmonics as follows
\begin{eqnarray}
&&\quad X^{\ell m}_{A}= \varepsilon_{A}^{\ B}Y^{\ell m}_{|B}, \quad 
U^{\ell m}_{AB}= \Omega_{AB}Y^{\ell m}\,,\nn\\
&&V^{\ell m}_{AB}=Y^{\ell m}_{|AB}+\frac{\ell(\ell+1)}{2}\Omega_{AB}Y^{\ell m}, \qquad W^{\ell
m}_{AB}= X^{\ell m}_{(A|B)}\nonumber
\end{eqnarray}
where $\Omega_{AB}=(1,\sin^2\theta)$, a bar denotes the covariant derivative with respect to the metric $\Omega_{AB}$,
and $\varepsilon_{AB}$ is the Levi-Civita tensor on the unit two-sphere. Hereafter capital roman indices run over the
angular coordinates ($\theta,\varphi$), while lower-case roman indices run over $t$ and $r$. We work in the frequency
domain and all the quantities are intended as Fourier transforms of some time-dependent quantity, i.e. schematically
$A(\omega,r)=\frac{1}{2\pi}\int dt \tilde{A}(t,r)e^{i\omega t}$. Finally we define $A'\equiv \partial
A(\omega,r)/\partial r$.
\subsection{Perturbation equations with a general source}
In our perturbative approach, the spacetime metric is $g_{\mu\nu}=g^{(0)}_{\mu\nu}+h_{\mu\nu}$ where $g^{(0)}_{\mu\nu}$
is the Schwarzschild metric (\ref{schwmet}), and $h_{\mu\nu}$ is the metric perturbation. We decompose $h_{\mu\nu}$ in
tensor spherical harmonics choosing the RW gauge, as in Ref.~\cite{Cardoso:2009pk} where the perturbation equations
without source were derived.

There are two families of perturbations, the odd (or axial) parity perturbations (in our gauge described by the
functions $h^{\ell m}_0(\omega,r)$ and $h^{\ell m}_1(\omega,r)$), and the even (or polar) parity perturbations (in our
gauge described by the functions $H^{\ell m}_0(\omega,r)$, $H^{\ell m}_1(\omega,r)$, $H^{\ell m}_2(\omega,r)$ and
$K^{\ell m}(\omega,r)$). Moreover, we decompose the scalar field as
\begin{equation}
\vartheta(t,r,\theta,\varphi)=\int_{-\infty}^\infty d\omega\,\sum_{\ell m}\frac{\Theta^{\ell
m}(\omega,r)}{r}Y^{\ell m}(\theta,\varphi) e^{-i\omega t}\,.\label{scalar_decomp}
\end{equation}
In the rest of the Appendix, we leave implicit the $\omega$ dependence and the $\ell$ and $m$ indices, a sum over which
is assumed.  The relevant first order Einstein equations for odd parity perturbations read
\begin{eqnarray}
\label{eq:oddfieldeqns}
 &&i\omega h_1' + h_0'' + \frac{2i\omega}{r} h_1 - \frac{2 (\lambda + 1) r - 4M}{r^3 f}  h_0+\nn\\
&&-\frac{96\pi M\alpha}{r^5}\left(r \eta-2\Theta\right)= P^t_{\omega \ell m}(r)\,,\label{E1} \\
&&-\omega^2 h_1 +i\omega h_0' - \frac{2i\omega}{r} h_0 + \frac{2 \lambda f }{r^2}  h_1+\nn\\
&&-\frac{96i\pi\alpha\omega M}{r^4}\Theta= P^r_{\omega \ell m}(r)\,, \label{E2}\\
&&\frac{i\omega}{f}  h_0 + f  h_1' + \frac{2M}{r^2}  h_1 = P_{\omega \ell m}(r)\,,\label{E3}
\end{eqnarray}
where $\lambda=(\ell+2)(\ell-1)/2$ and the source terms are the Fourier transforms of
\begin{eqnarray}
P^a (t,r)
&&\equiv  \frac{16 \pi r^2}{\ell (\ell +1)}\int T^{aB} X_{B}^{*} \, d\Omega\,,\\
P (t,r)  &&\equiv  16 \pi r^4 \frac{(\ell - 2)!}{(\ell + 2)!} \int T^{AB} W^{*}_{AB}\, d\Omega\,.
\end{eqnarray}
The three equations above are not independent, due to the Bianchi identity
\begin{equation}
\pa_t P^t +  \partial_r{P^r} + \frac{2}{r} P^r - \frac{2 \lambda}{r^2} P = 0\,.
\end{equation}
Defining the RW function as (notice the sign difference with respect to Ref.~\cite{Cardoso:2009pk})
\begin{equation}
Q^{\ell m}(\omega,r)=-\frac{f(r)}{r}h_1^{\ell m}(\omega,r)\,,\label{defRW}
\end{equation}
from Eqs.~\eqref{E1}-\eqref{E3} we obtain the equation for axial parity perturbations~\eqref{eqRW}, with source term
given by
\begin{eqnarray}
S_{RW}(\omega,r)&=&\frac{1}{2\pi}\int dt\,S_{RW}(t,r) e^{i\omega t}\,,\label{sourceFT_RW}\\
S_{RW}(t,r)&=&\frac{f}{r}\left[\frac{2}{r}\left(1-\frac{3M}{r}\right)P-f\pa_r
P+P^r\right]\,.\nonumber\\\label{sourceRW}
\end{eqnarray}
The perturbation equation for the scalar field can be computed replacing Eq.~\eqref{scalar_decomp} into
Eq.~\eqref{eqE2} and linearizing it. Using Eq.~\eqref{E2}, we obtain Eq.~\eqref{eqScalar} with source term given by
\begin{eqnarray}
S_{S}(\omega,r)&=&\frac{1}{2\pi}\int dt\,S_{S}(t,r) e^{i\omega t}\,,\label{sourceFT_Scalar}\\
S_{S}(t,r)&=&-if(r)\frac{6\ell(\ell+1)\alpha M}{r^4\beta\omega}P^r\,.\label{sourceScalar}
\end{eqnarray}
We note that this source term is entirely due to the CS coupling.  The polar sector, instead, is unaffected by the CS
coupling. Hence, the equation for polar perturbations~\eqref{eqZer} is the same as in general relativity with source
term given by~\cite{Martel:2003jj}:
\begin{widetext}
\begin{eqnarray}
S_{Z}(\omega,r)&=&\frac{1}{2\pi}\int dt\,S_{Z}(t,r) e^{i\omega t}\,,\label{sourceFT_Zer}\\
S_{Z}(t,r)&=&\frac{1}{(\lambda+1)\Lambda}\, \Biggl\{ 
	    r^2 f \biggl( f^2 \frac{\partial}{\partial r} N^{tt} - \frac{\partial}{\partial r}
N^{rr} \biggr) 
	    + r(\Lambda - f) N^{rr} + rf^2 N^\flat \nonumber \\
	   & & - \frac{f^2}{r\Lambda} \Bigl[ \lambda(\lambda-1)r^2 + (4\lambda-9)Mr + 15M^2
	    \Bigr] N^{tt} \Biggr\} + \frac{2f}{\Lambda}\, N^{r} - \frac{f}{r}\, N^\sharp
\,,\label{sourceZer}
\end{eqnarray}
where $\Lambda=\lambda+3M/r$ and we have defined
\begin{eqnarray}
&&N^{ab} = 8\pi \int T^{ab} Y^{\ell m*}\, d\Omega, \qquad
N^a = \frac{16\pi r^2}{\ell(\ell+1)}\, \int T^{aA} Y^{\ell m*}_{|A}\, d\Omega,\\
&&N^\flat = 8\pi r^2 \int T^{AB} U^{\ell m*}_{AB}\, d\Omega, \qquad
N^\sharp = \frac{32\pi r^4}{(\ell-1)\ell(\ell+1)(\ell+2)}\, \int T^{AB}
V^{\ell m*}_{AB}\, d\Omega,\,.
\end{eqnarray}
\end{widetext}
\subsection{Source term for point particle on geodesics}
Here we work out the explicit source terms for a point particle moving on geodesics around a Schwarzschild BH (see
e.g. Ref.~\cite{Martel:2003jj}). Let us consider a point-like particle on a timelike geodesic with coordinates
$z^{\mu}_{p}(\tau)$. The stress-energy tensor (in the time domain) is
\begin{equation}
T^{\mu\nu} = \mu \int \frac{d\tau}{\sqrt{-g}}\, u^{\mu}u^{\nu}\
\delta^{4}\left(x^{\alpha}-z^{\alpha}_{p}\right)\,,
\end{equation}
where $\mu$ is the mass of the particle, $\delta^{4}(x^{\alpha}-z_p^{\alpha})$ is the four-dimensional
Dirac delta, $\tau$ is the proper time and $u^{\mu}(\tau)=\dot{z}^{\mu}_{p}(\tau)$ is the
four-velocity. 
The integral above can be explicitly computed:
\begin{eqnarray}
T^{\mu\nu} =\mu\frac{u^\mu(t)u^\nu(t)}{r_p(t)^2
u^t(t)}\delta(r-r_p(t))\delta(\cos\theta)\delta(\varphi-\varphi_p(t))\,.\nn
\end{eqnarray}
We introduce the semi-latus rectum $p$ and the eccentricity $e$, as orbital parameters.  They are defined so that the
periastron and apastron are at $r=pM/(1+e)$ and $r=pM/(1-e)$, respectively.  In terms of these parameters, the energy
and angular momentum per unit mass of a point particle are
\begin{eqnarray}
\tilde{E}^2&=&\frac{(p-2-2e)(p-2+2e)}{p(p-3-e^2)}\,, \nonumber \\
\tilde{L}^2&=&\frac{M^2p^2}{p-3-e^2}\,,
\end{eqnarray}
and the four velocity reads
\begin{equation}
u^\mu=\left\{\frac{\tilde{E}}{f},\sqrt{\tilde{E}^2-\tilde{V}^2},0,\frac{\tilde{L}}{r^2}\right\}\,,
\end{equation}
where $\tilde{V}^2=f\left(1+\tilde{L}^2/r^2\right)$.

For point-like particles on geodesics the time-dependent source terms \eqref{sourceRW} and \eqref{sourceZer} can be
computed explicitly. They read \cite{Martel:2003jj}
\begin{eqnarray}
S_{RW}(t,r)&=&G_{RW}(t,r)\delta[r-r_{p}(t)]+\nn\\
&&+F_{RW}(t,r)\delta '[r-r_p(t)]\,,\label{sourceRW_fin}\\
S_{Z}(t,r)&=&G_{Z}(t,r)\delta[r-r_{p}(t)]+\nn\\
&&+F_{Z}(t,r)\delta '[r-r_p(t)]\,,\label{sourceZer_fin}
\end{eqnarray}
with
\begin{eqnarray}
G_{RW}(t,r) &=&\frac{f^{2}}{r^3}\Biggl[\frac{4}{r}\left(1-\frac{3M}{r}\right)A +B\Biggr], \nonumber
\\
F_{RW}(t,r) &=& -A\frac{f^3}{r^3},\nn\\
G_Z(t,r)&=& a \ Y^{*}(t) + b \ Y^{*}_{|\varphi}(t) + c \ U^{*}_{\varphi \varphi}(t)+ d \
V^{*}_{\varphi \varphi}(t)\,,\nonumber\\
&& \label{Gz(t,r)}\\
F_Z(t,r)&=& \frac{8\pi}{\lambda+1}\frac{f^2}{\Lambda}\frac{\tilde{V}^2}{\tilde{E}}\ Y^{*}(t),\nn
\end{eqnarray}
where 
\begin{eqnarray}
A &=&16 \pi \frac{(\ell-2)!}{(\ell+2)!} \frac{\tilde{L}^2}{\tilde{E}}\ W^{*\ell m}_{\varphi
\varphi}(t),\nn\\
 B &=& \frac{8 \pi}{\lambda+1} \frac{\tilde{L}}{\tilde{E}}\ u^{r}\ X^{*\ell
m}_{\varphi}(t)\,,\nn\\
a &=& \frac{8\pi}{\lambda+1}\frac{f^2}{r\Lambda^2}\Biggl\{
\frac{6M}{r}\tilde{E}+\nn\\
&&-\frac{\Lambda}{\tilde{E}}\biggl[\lambda+1-\frac{3M}{r}+\frac{\tilde{L}^2}{r^2}
\left(\lambda+3-\frac{7M}{r}\right)\biggr]\Biggr\}\,,\nn\\
b &=& \frac{16 \pi}{\lambda+1} \frac{\tilde{L}}{\tilde{E}}\frac{f^2}{r^2\Lambda}\ u^{r}, \qquad
c = \frac{8\pi}{\lambda+1} \frac{\tilde{L}^2}{\tilde{E}}\frac{f^3}{r^3 \Lambda},\,,\nn\\
d &=& -32 \pi \frac{(\ell-2)!}{(\ell+2)!} \frac{\tilde{L}^2}{\tilde{E}}\frac{f^2}{r^3}.
\end{eqnarray}
In the above expressions, $Y(t)$, $X_A(t)$, $U_{AB}(t)$, $V_{AB}(t)$, $W_{AB}(t)$ denote the scalar, vector and tensor
spherical harmonics evaluated at the angular position of the particle $\varphi_{p}(t)$; thus, $Y(t)$ is shorthand
notation for $Y^{\ell m}(\pi/2,\varphi_{p}(t))$, and the same holds for the other harmonics.

Since the orbital motion takes place in the equatorial plane, each spherical harmonic function is evaluated at
$\theta_{p}=\pi/2$.  An important consequence of this is that the source term for the Zerilli function vanishes when
$\ell+m$ is odd, while the source term for the RW function vanishes when $\ell+m$ is even.

Finally, the source term for the scalar equation \eqref{sourceScalar} is
\begin{equation}
S_S(t,r)= G_{S}(t,r)\delta[r-r_{p}(t)]\,,\label{sourceScalar_fin}
\end{equation}
where
\begin{equation}
G_S(t,r)=-if^2\frac{96\pi\alpha
M}{r^4\beta\omega}\frac{\tilde{L}}{\tilde{E}}\frac{\sqrt{\tilde{E}^2-\tilde{V}^2}}{r_p^2(t)}
X_\varphi^*(t)\,.
\end{equation}
Notice that $S_S$, at variance with $S_{RW}$ and $S_{Z}$, does not contain derivatives of the Dirac delta.  Note also
that if the orbit is circular, $\tilde E=\tilde V$, thus $S_S=0$. This can be traced back to Eq.~\eqref{sourceScalar},
and to the fact that $P^r\sim T^{r\mu}\sim u^r=0$ for circular orbits.

\subsection{Source describing a particle in circular orbit}
All quantities in the previous section are considered in the frequency domain: they depend on $r$ and $\omega$, although
the dependence on $\omega$ has often been left implicit. To compute the sources $S_Z(\omega,r)$ and $S_{RW}(\omega,r)$,
one should first consider the time domain sources $S_Z(t,r)$, $S_{RW}(t,r)$, and then compute their Fourier
transforms. This operation proceeds straightforwardly in the case of a circular orbit $r\equiv\bar r$. Indeed, we have
\begin{eqnarray} r_p(t)=\bar r\,,\qquad \phi_p(t)=\omega_Kt\,,\nn \end{eqnarray} where the Keplerian frequency reads
\begin{equation}
\omega_K=\sqrt{\frac{M}{r^3}}\,.\label{kepl}
\end{equation}
Furthermore the geodesics energy, angular momentum and four-velocity respectively read
\begin{eqnarray}
\tilde E&=&\frac{r-2M}{\sqrt{r(r-3M)}}\nonumber\\
\tilde L&=&r\sqrt{\frac{M}{r-3M}}\nonumber\\
u^\mu&=&\left(\sqrt{\frac{r}{r-3M}},0,0,\frac{1}{r}\sqrt{\frac{M}{r-3M}}\right)\,.
\end{eqnarray}
Using the definitions above, the source terms \eqref{sourceRW_fin}, \eqref{sourceZer_fin}, \eqref{sourceScalar_fin}
reduce to
\begin{eqnarray} 
S_{RW}(t,r)&=&G_{RW}(t,r)\delta(r-\bar r)+F_{RW}(t,r)\delta'(r-\bar r)\nonumber\\ 
S_{Z}(t,r)&=&G_{Z}(t,r)\delta(r-\bar r)+F_{Z}(t,r)\delta'(r-\bar r)\nonumber\\ 
S_S(t,r)&=&0\,.
\end{eqnarray} 
The dependence on $t$ comes from the tensor spherical harmonics. For instance Eq.~\eqref{Gz(t,r)} now reads (hereafter
we write explicitly the indexes $\ell,m$)
\begin{equation}
G^{\ell m}_Z(t,r)=aY^{*\ell m}(\theta_p(t),\phi_p(t))+\dots
\end{equation}
with
\begin{eqnarray}
Y^{*\ell m}(\theta_p(t),\phi_p(t)) &&=Y^{*\ell m}\left(\frac{\pi}{2},0\right)e^{-\ii
m\phi_p(t)}\nn\\
&&=Y^{*\ell m}\left(\frac{\pi}{2},0\right) e^{-\ii m\omega_Kt}\nn\,.
\end{eqnarray}
and similarly for the other terms in Eq.~\eqref{Gz(t,r)}.  Hence, we can write $G^{\ell m}_Z(t,r)$ as a quantity which
does not depend on $t$ (conventionally we will indicate it with a hat) times $e^{-\ii m\omega_Kt}$:
\begin{equation}
G^{\ell m}_Z(t,r)=\hat G^{\ell m}_Z(r) e^{-\ii m\omega_Kt}
\end{equation}
where 
\begin{equation}
\hat
G^{\ell m}_Z(r)=aY^{*\ell m}\left(\frac{\pi}{2},0\right)+\dots\,,
\end{equation}
The same can be done with all other quantities.  Then, the Fourier transform gives:
\begin{eqnarray}
G^{\ell m}_Z(\omega,r)&&=\frac{1}{2\pi}\int_{-\infty}^{+\infty}dt\hat G^{\ell m}_Z(r)e^{-\ii m\omega_K
t}e^{\ii\omega t}\nn\\
&&=
\hat G^{\ell m}_Z(r)\delta(\omega-m\omega_K)\,,\nn
\end{eqnarray}
and the same holds for the other quantities. Therefore,
\begin{eqnarray}
S^{\ell m}_{RW}(\omega,r)&=&\delta(\omega-m\omega_K)\times\nn\\
&&\left[\hat G^{\ell m}_{RW}(r)\delta(r-\bar r)+\hat
F^{\ell m}_{RW}(r)\delta'(r-\bar r)\right]\,,\nonumber\\
&&\label{SRWomega}\\
S^{\ell m}_{Z}(\omega,r)&=&\delta(\omega-m\omega_K)\times\nn\\
&&\left[\hat G^{\ell m}_{Z}(r)\delta(r-\bar r)+\hat
F^{\ell m}_{Z}(r)\delta'(r-\bar r)\right]\,.\nonumber\\
&&\label{SZomega}
\end{eqnarray}

\section{Perturbative Green's function approach}\label{app:green}
In this Appendix we develop a perturbative Green's function approach to solve Eqs.~\eqref{eqRW}-\eqref{eqZer}. This
approach is valid in the small coupling limit. We shall compare it with the general method (valid for any coupling)
adopted in the main text. The perturbative approach can be useful for possible analytical calculations and it is also
important as independent check for numerical results.

As long as the dimensionless CS coupling $\zeta$ (\ref{zetadef}) is small, Eqs.~\eqref{eqRW}-\eqref{eqZer} can be solved
by a perturbative scheme. Indeed, as we shall see, the Regge-Wheeler function and the scalar field are consistent with
the ansatz
\begin{eqnarray}
Q&=&Q^{(0)}+\zeta Q^{(1)}\,,\nonumber\\
\Theta&=&\Theta^{(0)}+\zeta \Theta^{(1)}\,.\label{Greenperturbative}
\end{eqnarray}
The function $Q^{(0)}$ is the solution of the standard Regge-Wheeler equation in general relativity
\begin{eqnarray}
\left[\frac{d^2}{d r_*^2}\!+\!\omega^2\!-\!V_{RW}(r)\right]\!Q(r)&=&S_{RW}(r)\,,\nonumber\\
\end{eqnarray}
and can be solved by the Green's function approach, as in Section~\ref{greeneven}, by considering the particle
source only:
\begin{equation}
Q^{(0)}(r)=\frac{Q_+(r)}{W_{RW}}\!\!\int_{-\infty}^r\!\!\!\!\!\!dr_*{Q_-S_{RW}}\!+\!
\frac{Q_-(r)}{W_{RW}}\!\!\int^{+\infty}
_r\!\!\!\!\!\!dr_*{Q_+S_{RW}}
\end{equation}
(note that the Green's function expression is linear in the source).  The function $\Theta^{(0)}$ is the lowest order
contribution (in the CS coupling) to the scalar field $\Theta$.  Its equation can be found replacing the ansatz
(\ref{Greenperturbative}) in Eq.~(\ref{sceq}): at lowest order in $\zeta$, it gives
\begin{equation}
\left[\frac{d^2}{d r_*^2}+\omega^2-V_S\right]\Theta^{(0)}=S_S
-f\frac{(\ell+2)!}{(\ell-2)!}\frac{6\ii M\alpha}{\omega r^5\beta}Q^{(0)}
\label{sceq1}\,,
\end{equation}
where $S_S\sim\alpha/\beta$ is given in Eq.~\eqref{sourceScalar} and $V_S$ is Eq.~\eqref{potentialScalar} at order
$\alpha$.  Defining an effective source term
\begin{equation}
\bar S_S(r)\equiv S_S -f\frac{(\ell+2)!}{(\ell-2)!}\frac{6\ii M\alpha}{\omega
r^5\beta}Q^{(0)}(r)\,,\label{barss}
\end{equation}
we can write the solution of (\ref{sceq1}) as 
\begin{eqnarray}
\Theta^{(0)}(r)&=&\frac{1}{W_\Theta}\left[\Theta_+(r)\int_{-\infty}^rdr_*
{\Theta_-\bar S_S}\right.\nonumber\\
&&\left.+\Theta_-(r)\int^{+\infty}_rdr_*{\Theta_+\bar S_S}\right]\,.
\end{eqnarray}
Finally, if we denote the scalar field source in the Regge-Wheeler equation (at lowest order in $\zeta$) as
\begin{equation}
\bar S_{RW}(r)\equiv \frac{96\ii\pi M\omega f}{r^5}\alpha\Theta^{(0)}(r)\,,\label{barsq}
\end{equation}
we get that the main correction to the Regge-Wheeler function is
\begin{eqnarray}
\zeta Q^{(1)}&&=\frac{1}{W_{RW}}\left[Q_+(r)\int_{-\infty}^rdr_*{Q_-\bar
S_{RW}}\right.\nn\\
&&\left.+Q_-(r)\int^{+\infty}_rdr_*{Q_+\bar S_{RW}}
\right]\,.\label{delq}
\end{eqnarray}
Note that $\bar S_S\sim \alpha/\beta$, therefore $\Theta^{(0)}\sim\alpha/\beta$, and then
$S_{RW}\sim\alpha\Theta^{(0)}\sim\zeta$, 
consistently with Eq.~\eqref{delq}.

At infinity and at the horizon, where the fluxes are computed, we get
\begin{eqnarray}
Q(r_*\rightarrow\pm\infty)&&=(Q^{(0)}+\zeta Q^{(1)})(r_*\rightarrow\pm\infty)\nn\\
&&=\frac{e^{\pm\ii\omega
r_*}}{W_{RW}}\int_{-\infty}^\infty dr_*{Q_\mp(S_{RW}+\bar S_{RW})}\,.\nn\\
\end{eqnarray}
Even for circular orbits, $\bar S_{RW}\sim\Theta^{(0)}$ is defined throughout the entire spacetime, thus the evaluation
of the integrals above is more involved than in general relativity, in which we simply have $S_{RW}\sim\delta(r-\bar r)$
for a circular orbit. This makes the perturbative approach much more time-consuming than the non-perturbative approach.

It is straightforward to extract the quantities relevant to compute the fluxes, Eqs.~\eqref{flux_grav} and \eqref{flux_scalar}. 
Indeed, by applying the same procedure explained in the main text, we
can define ${\cal Q}^{(0)\ell m}_\pm(\omega,r)=\bar {\cal Q}^{(0)\ell   m}_\pm(r)\delta(\omega-m\omega_K)e^{\pm i\omega r_*}$ with
\begin{equation}
\bar {\cal Q}_\pm^{(0)\ell m}(r)=\frac{Q_\mp(r)}{W_{RW}}\left[\frac{Q_\pm\hat G^{\ell m}_{RW}}{f}-\left(\frac{Q_\pm\hat
F^{\ell m}_{RW}}{f}\right)'\right]_{\bar r}\,,\nn
\end{equation}
for $r\lessgtr\bar r$ respectively.  Then, from Eqs.~(\ref{barss})-(\ref{delq}) one can compute the quantities
${\varTheta}^{(0)\ell m}_\pm(r)$, $\bar S^{\ell m}_{RW}(r)$, ${\cal Q}^{(1)\ell m}_\pm(r)$, and finally the complete
Regge-Wheeler function ${\cal Q}^{\ell m}_\pm={\cal Q}^{(0)\ell m}_\pm+\zeta {\cal Q}^{(1)\ell m}_\pm$ at infinity,
which has the form
\begin{equation}
{\cal Q}^{\ell m}_\pm(\omega)=\bar {\cal Q}^{\ell m}_\pm\delta(\omega-m\omega_K)e^{\pm\ii\omega r_*}\,.\label{QbarQ2}
\end{equation}
In the same way, the scalar field perturbation $\varTheta^{(0)\ell m}_\pm(\omega)\equiv
\Theta^{(0)\ell m}_\pm(\omega,r_*\rightarrow\pm\infty)$ has the form
\begin{equation} 
\varTheta^{(0)\ell m}_\pm(\omega)=\bar \varTheta^{(0)\ell m}_\pm\delta(\omega-m\omega_K)e^{\pm\ii\omega r_*}\,.\label{ThbarTh2}
\end{equation} 
Notice that $\Theta^{(0)}\sim\alpha/\beta$, so that the corresponding energy flux is ${\cal O}(\zeta)$, consistently
with our expansion~\eqref{delq}. The contribution of $\zeta\Theta^{(1)}$ to the energy flux is ${\cal O}(\zeta^2)$ and
it is neglected in this approximation.
\subsubsection{Comparison between the perturbative and the non-perturbative approach}
\begin{figure*}[htb]
\begin{center}
\begin{tabular}{cc}
\epsfig{file=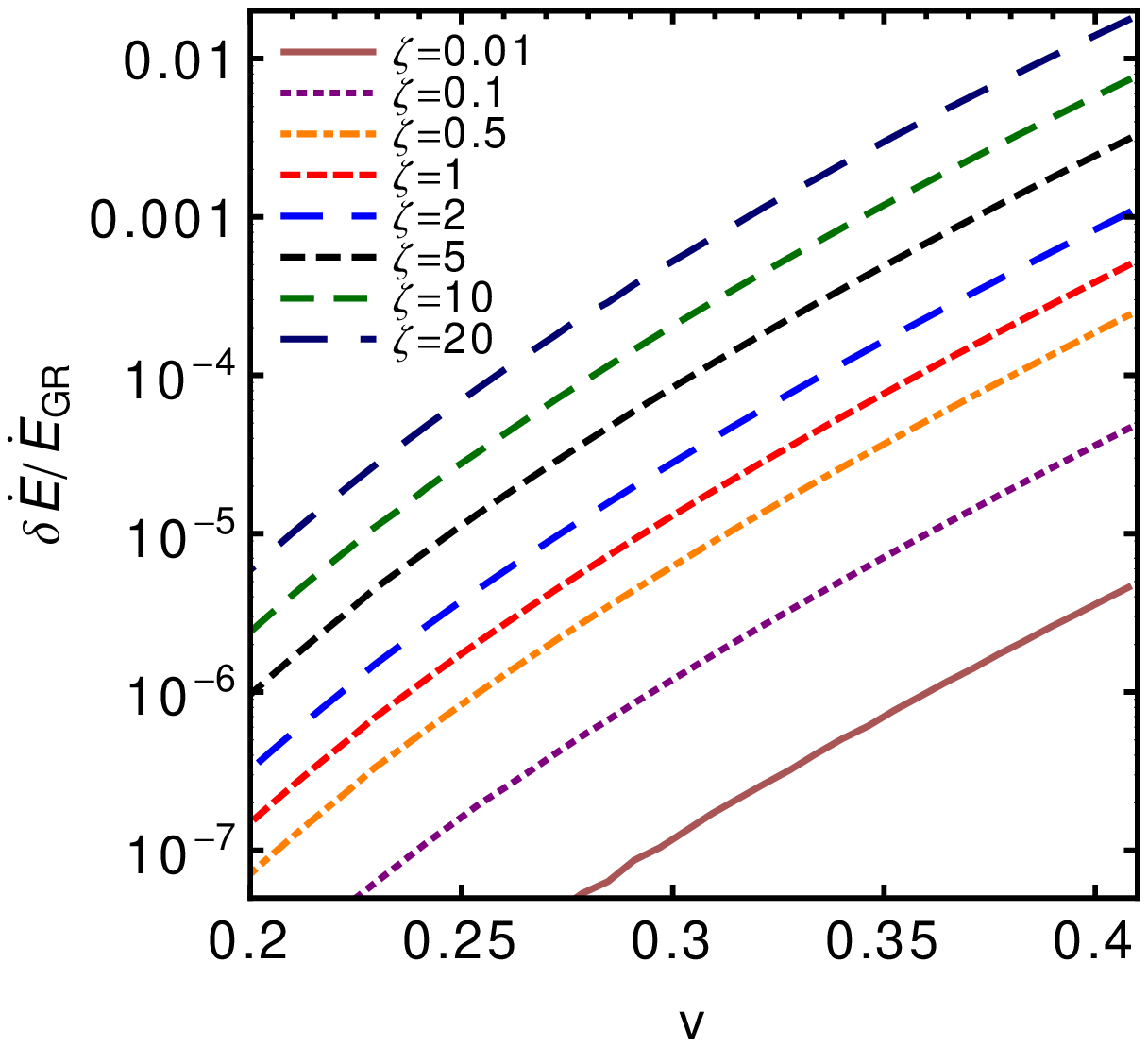,height=6.5cm,angle=0}&
\epsfig{file=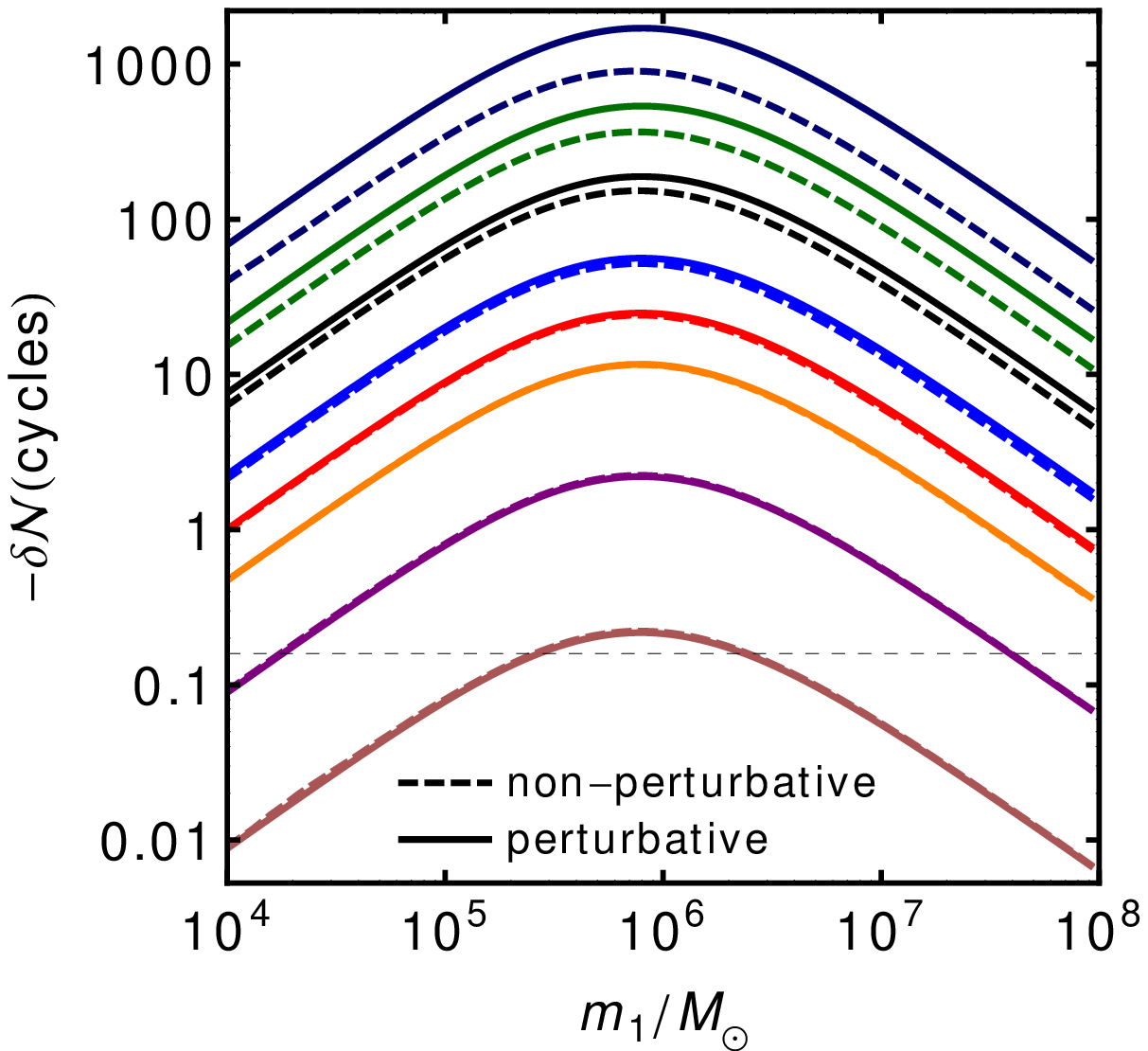,height=6.5cm,angle=0}
\end{tabular}
\caption{(Color online) Comparison between the non-perturbative method
  described in the main text and the perturbative method described in
  Appendix~\ref{app:green}. Left: Corrections to the total flux
  emitted. Right: Corrections to the number of cycles for $m_2=1.4
  M_\odot$. In both panels, from below to top:
  $\zeta=0.01,0.1,0.5,1,2,5,10,20$. For $\zeta\gtrsim1$ the 
perturbative method overestimates the total flux, resulting in a larger (in absolute value) correction to $\delta{\cal N}$.
  \label{fig:comparison_NPVSP}}
\end{center}
\end{figure*}

Comparing the perturbative method with the general approach discussed in the main text is important for two
reasons. First, it gives an independent check of both methods. Second, it allows to assert the validity region of the
perturbative approach. In Fig.~\ref{fig:comparison_NPVSP} we compare the corrections to the total energy flux and to the
number of cycles obtained by the two methods. 

When $\zeta\ll1$, the methods agree very well, confirming each
other. 
For example, when $p\sim6$, the emitted fluxes computed with the two methods agree within $0.1\%$ 
or better for $\zeta\lesssim0.3$, whereas they differ by $\sim(1\%,5\%,10\%,50\%,100\%)$ when $\zeta\sim(0.5,1,2,10,20)$, respectively. 
Remarkably, the perturbative approach is valid within a few percent up to $\zeta\sim1$ whereas, for larger values of $\zeta$, it gives
an \emph{overestimated} flux.
%
\subsubsection{On the dominance of fluxes at the horizon in DCS theory}
%
\begin{figure}[htb]
\begin{center}
\begin{tabular}{cc}
\epsfig{file=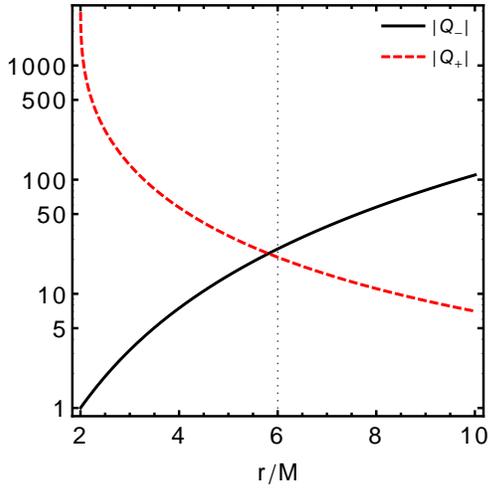,height=6.5cm,angle=0}&
\end{tabular}
\caption{(Color online) Homogeneous solutions of the Regge-Wheeler equation, $|Q_{\pm}|$. 
\label{fig:comphom}}
\end{center}
\end{figure}
One of the curious results borne out of this study, with important consequences for the main observational prospects,
concerns the fluxes at the horizon: while in general relativity fluxes at the horizon are orders of magnitude smaller
than at infinity, this does not occur in DCS gravity; as summarized in Eqs.~\eqref{pnresult} the {\it corrections}
imparted by the DCS coupling affect more strongly the flux at the horizon rather than at infinity. This can be
understood from our perturbative analysis in the following way.  From the above study, the axial Regge-Wheeler function
in general relativity behaves as
\begin{equation}
\left|Q^{(0)}(r_*\to \pm\infty)\right|=\left|\frac{1}{W_{RW}}\int_{-\infty}^{+\infty}dr_* Q_{\mp}S_{RW}\right|\,.
\end{equation}
We assume for simplicity circular orbits and that $S_{RW}\sim \delta(r-\bar r)$, i.e. we neglect the $\delta'(r - \bar r)$ contribution; however,
this discussion can be easily generalized to non-circular orbits.  The fluxes at the horizon and at infinity then depend
on the value of the homogeneous solutions at $r = \bar r$. From Fig.~\ref{fig:comphom} where we plot these homogeneous
solutions, we see that, for $r > \bar r\geq 6M$,
\begin{equation}
\left|Q_+\right|^2>\left|Q_-\right|^2\,,
\end{equation}
which explains why the flux at the horizon is smaller than that at infinity. On the other hand, in this perturbative
approach, the DCS correction to the Regge-Wheeler function behaves as
\begin{equation}
\left|\zeta Q^{(1)}(r_*\to \pm\infty)\right|=\left|\frac{1}{W_{RW}}\int_{-\infty}^{+\infty}dr_* Q_{\mp}\bar S_{RW}\right|\,.
\end{equation}
Since $\bar S_{RW}(r) \sim \Theta^{(0)}/r^{5}$, the largest contribution comes from the near-horizon region $2M<r<6M$,
where (see Fig~\ref{fig:comphom})
\begin{equation}
\left|Q_+\right|^2<\left|Q_-\right|^2\,,
\end{equation}
therefore the largest DCS correction to the flux is at the horizon.

Analogously, the DCS scalar field behaves as
\begin{equation}
\left|\varTheta^{(0)}(r_*\to \pm\infty)\right|=\left|\frac{1}{W_{RW}}\int_{-\infty}^{+\infty}dr_* \Theta_{\mp}\bar S_S(r)\right|\,.
\label{intth0}
\end{equation}
The homogeneous solutions of the scalar equation, $\Theta_\pm$, are similar to the Regge-Wheeler homogeneous solutions
$Q_{\pm}$; in particular, close to the horizon one has $\left|\Theta_+\right|^2<\left|\Theta_-\right|^2$, and $\bar
S_{S}(r) \sim Q^{(0)}/r^{5}$, therefore the DCS scalar flux is larger at the horizon than at infinity.

\section{Comparison of different ways to estimate the frequency one year prior to merger}\label{app:comparison}
%
\begin{figure*}[htb]
\begin{center}
\begin{tabular}{cc}
\epsfig{file=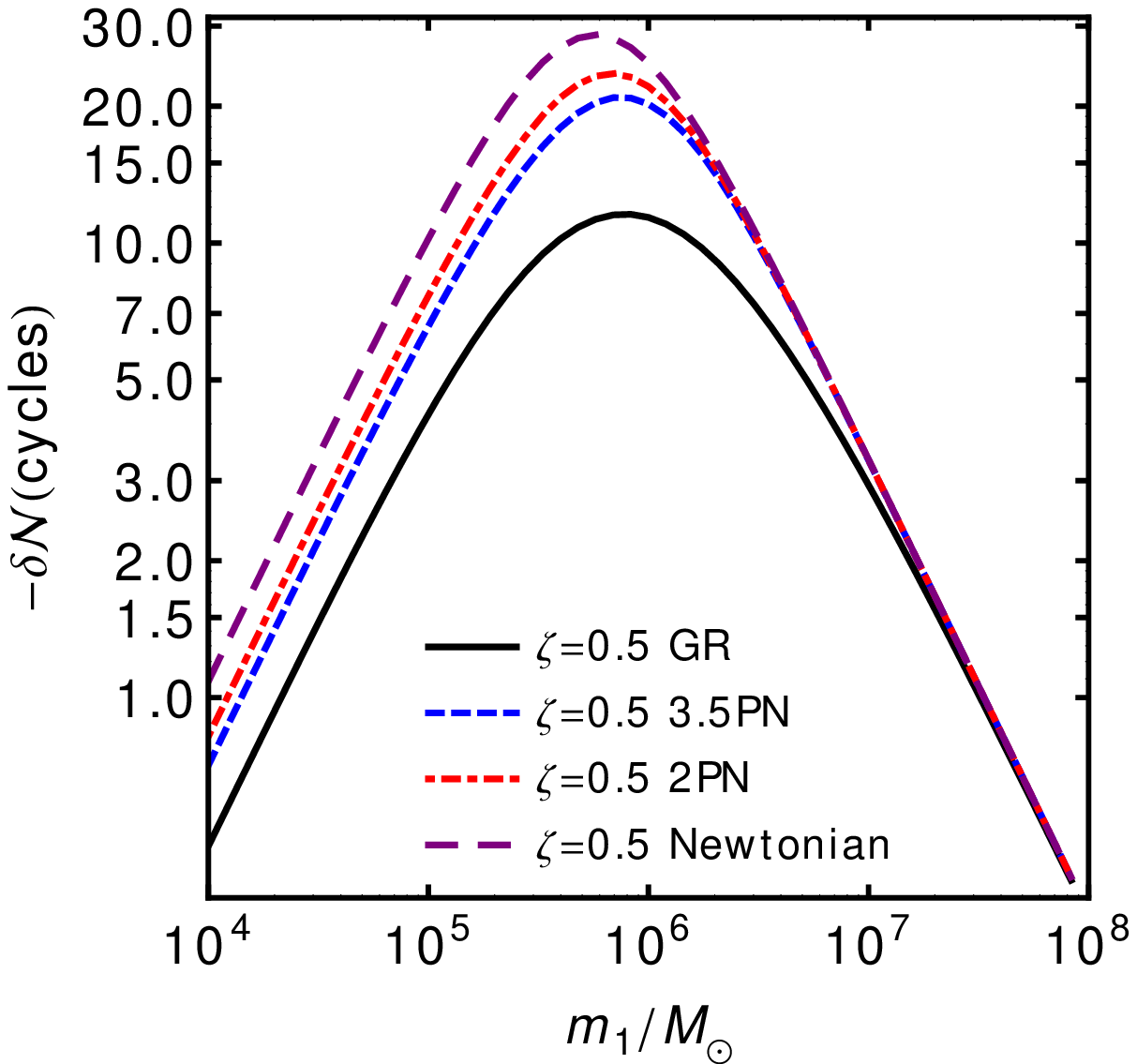,height=6.5cm,angle=0}&
\epsfig{file=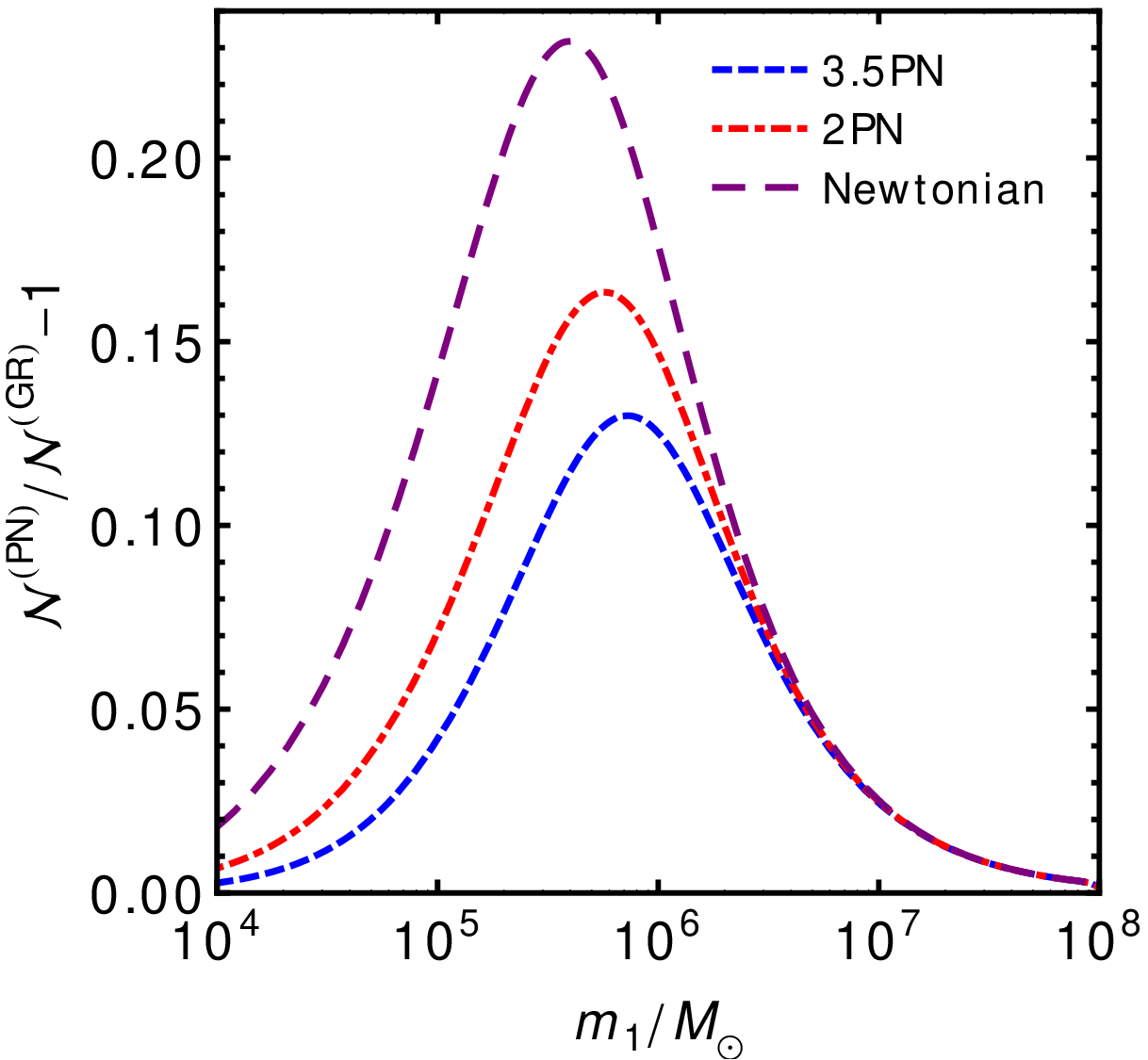,height=6.5cm,angle=0}
\end{tabular}
\caption{(Color online) Left: Same as right panel of Fig.~\ref{fig:deltaN}, but with different prescriptions to
  compute the frequency one year prior to merger.
The first prescription (``GR'') is adopted in the main text and takes circular geodesic motion and the energy fluxes 
numerically computed in general relativity for a particle in circular orbit. The second (``3.5PN'') and third (``2PN'') 
prescriptions take respectively the 3.5PN formula~\cite{Blanchet:2004ek,Buonanno:2006ui} and the 2PN formula, 
Eq.~(\ref{fdot2PN}) for the evolution of the frequency, while the third (``Newtonian'') is its truncation at lowest
order. Right: relative difference between the total number of cycles in different approximations with respect to our 
geodesic-based approach for $\zeta=0$ (general relativity). At low frequencies all prescriptions yield identical results, 
while at higher frequencies (smaller central masses), they can differ by a factor of order two.\label{fig:comparison}}
\end{center}
\end{figure*}
In our calculation of the evolution of orbital frequency with time, we have adopted an adiabatic approximation, where
the fluxes are determined numerically from those of a particle in geodesic motion. In the literature one sometimes finds
other alternative calculations, with which we now compare our (more accurate) results.
 
One of the most common alternatives for estimating $\dot f$ in Eqs.~\eqref{cycles} and~\eqref{time} consists in
taking a PN approximation which, at 2PN level, reads~\cite{Blanchet:1995ez}:
\begin{eqnarray}
\dot f&&=\frac{96}{5\pi}\eta M^{5/3}(\pi f)^{11/3}\left[1-\left(\frac{743}{336}+\frac{11}{4}\eta\right)(\pi M f)^{2/3}+\right.\nn\\
&&\left.4\pi^2 M f+\left(\frac{34103}{18144}+\frac{13661}{2016}\eta+\frac{59}{18}\eta^2\right)(\pi M f)^{4/3}\right]\,,\nn\\\label{fdot2PN}
\end{eqnarray}
(see e.g. Refs.~\cite{Blanchet:2004ek,Buonanno:2006ui} for the 3.5PN formula). In the equation above, $M=m_1+m_2$ is the
total mass of the two-body system and $\eta=m_1m_2/M^2=\mu/M$. Neglecting 2PN terms in Eq.~\eqref{fdot2PN}, we can solve
Eq.~\eqref{time} for $f_\text{1yr}$ analytically:
\begin{equation}
f_\text{1yr}=\frac{5^{3/8}}{2\sqrt{2}\pi M}\left(405+16\eta\frac{T_\text{obs}}{M}\right)^{-3/8}\,.
\end{equation}
For example, from the formula above $f_\text{1yr}\sim0.00274$~Hz for $T_\text{obs}=1$yr, $m_1=10^6 M_\odot$ and
$m_2=10M_\odot$, while, including 2PN corrections, for the same parameters we obtain $f_\text{1yr}\sim0.00252$~Hz, and
including 3.5PN corrections~\cite{Blanchet:2004ek,Buonanno:2006ui} we get $f_\text{1yr}\sim0.00242$~Hz. Our own
geodesic-based approach plus the numerical fluxes yields $f_\text{1yr}\sim0.00223$~Hz. Notice also that, if we insert
the quadrupole formula for the energy flux of a particle in a circular orbit,
\begin{equation}
\dot E\equiv\dot E_N=\frac{32}{5}\frac{\mu^2 M^3}{{\bar r}^5}\,,
\end{equation}
in Eq.~\eqref{dotfFIN} and expand for ${\bar r}/M\gg1$, then we recover the PN formula~\eqref{fdot2PN} at first order. 

The errors introduced by the PN approximation may be evaluated with a hybrid approach as follows. We compute the
modification $\delta {\cal N}$ using the numerical flux $\dot E_\text{DCS}$ in Eq.~\eqref{correctionN} but, in order to estimate
the lower boundary of the integral~\eqref{time}, we use four different prescriptions: (i) $\dot f$ is computed
numerically within our geodesic-based approach, as explained in the main text; (ii) $\dot f$ is given by its 3.5PN
expansion (see Eq.~(32) in Ref.~\cite{Buonanno:2006ui}); (iii) $\dot f$ is given by its 2PN expansion~(\ref{fdot2PN});
(iv) $\dot f$ is given by the truncation of Eq.~(\ref{fdot2PN}) at Newtonian order.  These different prescriptions
affect the value of $f_\text{in}$ and, in turn, Eq.~\eqref{correctionN}.

This is shown in the left panel of Fig.~\ref{fig:comparison}. For large central masses, i.e. low frequencies, all the
different prescriptions yield basically the same result to a good accuracy.  For smaller central masses, we expect
relativistic effects to become important: the system enters the LISA band when the small mass is already close to the ISCO, and where the PN expansion is less accurate. Indeed, in this regime the PN formula can give factors of order $\sim 2$ difference with respect to the more accurate prescription adopted in the main text.

As shown in the right panel of Fig.~\ref{fig:comparison}, similar deviations are observed also in general
relativity. Indeed, if we compute ${\cal N}$ from Eq.~\eqref{cycles} using different prescriptions for $f_\text{in}$, we
find that the number of cycles computed using the PN formula may be \emph{overestimated} by $\sim 20\%$.

As discussed in Section \ref{dcycles}, the method based on the adiabatic approximation is better suited to deal with
EMRIs.  We note here that this is true not only in alternative theories like DCS gravity, but also in general
relativity.
\section{Expressions for the coefficients $C_\pm^{(i)}$ and $D_\pm^{(i)}$}\label{app:coeff}
The coefficients in Eqs.~\eqref{solQ} and \eqref{solTheta} depend
solely on the solutions of the homogeneous system associated to Eq.~\eqref{system} and their explicit forms read
\begin{widetext}
 \begin{eqnarray}
 C_+^{(1)}&=&\Delta^{-1}\left[\Theta_-^{(2)}Q_-^{(1)}
   {\Theta'}_+^{(2)}-\Theta_+^{(2)}Q_-^{(1)} {\Theta'}_-^{(2)}-
\Theta_-^{(1)}Q_-^{(2)} {\Theta'}_+^{(2)}+\Theta_+^{(2)}
  Q_-^{(2)} {\Theta'}_-^{(1)}+\Theta_-^{(1)}Q_+^{(2)} {\Theta'}_-^{(2)}-\Theta_-^{(2)}Q_+^{(2)} {\Theta'}_-^{(1)}\right]\,,\nn\\
 C_+^{(2)}&=&\Delta^{-1}\left[-\Theta_-^{(2)}Q_-^{(1)}
   {\Theta'}_+^{(1)}+\Theta_+^{(1)}Q_-^{(1)} {\Theta'}_-^{(2)}+
\Theta_-^{(1)}Q_-^{(2)} {\Theta'}_+^{(1)}-\Theta_+^{(1)}
  Q_-^{(2)} {\Theta'}_-^{(1)}-\Theta_-^{(1)}Q_+^{(1)} {\Theta'}_-^{(2)}+\Theta_-^{(2)}Q_+^{(1)} {\Theta'}_-^{(1)}\right]\,,\nn\\
 C_-^{(1)}&=&\Delta^{-1}\left[\Theta_+^{(1)}Q_-^{(2)}
   {\Theta'}_+^{(2)}-\Theta_+^{(2)}Q_-^{(2)} {\Theta'}_+^{(1)}-
\Theta_-^{(2)}Q_+^{(1)} {\Theta'}_+^{(2)}+\Theta_+^{(2)}
  Q_+^{(1)} {\Theta'}_-^{(2)}+\Theta_-^{(2)}Q_+^{(2)} {\Theta'}_+^{(1)}-\Theta_+^{(1)}-Q_+^{(2)} {\Theta'}_-^{(2)}\right]\,,\nn\\
 C_-^{(2)}&=&\Delta^{-1}\left[-\Theta_+^{(1)}Q_-^{(1)}
   {\Theta'}_+^{(2)}+\Theta_+^{(2)}Q_-^{(1)} {\Theta'}_+^{(1)}+
\Theta_-^{(1)}Q_+^{(1)} {\Theta'}_+^{(2)}-\Theta_+^{(2)}
  Q_+^{(1)} {\Theta'}_-^{(1)}-\Theta_-^{(1)}Q_+^{(2)} {\Theta'}_+^{(1)}+\Theta_+^{(1)}Q_+^{(2)} {\Theta'}_-^{(1)}\right]\,,\label{defCC}
\end{eqnarray}
and
 \begin{eqnarray}
 D_+^{(1)}&=& \Delta^{-1}\left[-\Theta_+^{(2)} {Q'}_-^{(1)}
   Q_-^{(2)}+\Theta_-^{(2)} {Q'}_-^{(1)} Q_+^{(2)}+
\Theta_+^{(2)} {Q'}_-^{(2)} Q_-^{(1)}-\Theta_-^{(1)} {Q'}_-^{(2)}
   Q_+^{(2)}-\Theta_-^{(2)} {Q'}_+^{(2)} Q_-^{(1)}+\Theta_-^{(1)} {Q'}_+^{(2)} Q_-^{(2)}\right] \,,\nn\\
 D_+^{(2)}&=&\Delta^{-1}\left[ \Theta_+^{(1)} {Q'}_-^{(1)}
   Q_-^{(2)}-\Theta_-^{(2)} {Q'}_-^{(1)} Q_+^{(1)}-
\Theta_+^{(1)} {Q'}_-^{(2)} Q_-^{(1)}+\Theta_-^{(1)} {Q'}_-^{(2)}
   Q_+^{(1)}+\Theta_-^{(2)} {Q'}_+^{(1)} Q_-^{(1)}-\Theta_-^{(1)}{Q'}_+^{(1)} Q_-^{(2)}\right]\,,\nn\\
 D_-^{(1)}&=&\Delta^{-1}\left[ -\Theta_+^{(2)} {Q'}_-^{(2)}
   Q_+^{(1)}+\Theta_+^{(1)} {Q'}_-^{(2)} Q_+^{(2)}+
\Theta_+^{(2)} {Q'}_+^{(1)} Q_-^{(2)}-\Theta_-^{(2)} {Q'}_+^{(1)}
   Q_+^{(2)}-\Theta_+^{(1)} {Q'}_+^{(2)} Q_-^{(2)}+\Theta_-^{(2)} {Q'}_+^{(2)} Q_+^{(1)}\right]\,,\nn\\
 D_-^{(2)}&=&\Delta^{-1}\left[ \Theta_+^{(2)} {Q'}_-^{(1)}
   Q_+^{(1)}-\Theta_+^{(1)} {Q'}_-^{(1)} Q_+^{(2)}-
\Theta_+^{(2)} {Q'}_+^{(1)} Q_-^{(1)}+\Theta_-^{(1)} {Q'}_+^{(1)}
   Q_+^{(2)}+\Theta_+^{(1)} {Q'}_+^{(2)} Q_-^{(1)}-\Theta_-^{(1)}{Q'}_+^{(2)} Q_+^{(1)}\right]\,,\label{defDD}
\end{eqnarray}
where 
\begin{eqnarray}
 \Delta&=&{\Theta'}_+^{(2)} \left[\Theta_+^{(1)} {Q'}_-^{(1)}
   Q_-^{(2)}-\Theta_-^{(2)} {Q'}_-^{(1)} Q_+^{(1)}-
\Theta_+^{(1)} {Q'}_-^{(2)} Q_-^{(1)}+\Theta_-^{(1)}
   {Q'}_-^{(2)} Q_+^{(1)}\right]+\nn\\
&+&{Q'}_+^{(1)} \left[\Theta_-^{(2)} Q_-^{(1)} {\Theta'}_+^{(2)}-\Theta_+^{(2)} Q_-^{(1)} {\Theta'}_-^{(2)}-\Theta_-^{(1)}
   Q_-^{(2)} {\Theta'}_+^{(2)}+\Theta_+^{(2)} Q_-^{(2)} {\Theta'}_-^{(1)}+\Theta_-^{(1)} Q_+^{(2)} {\Theta'}_-^{(2)}-\Theta_-^{(2)} Q_+^{(2)}
   {\Theta'}_-^{(1)}\right]+\nn\\
&+&{Q'}_+^{(2)} \left[-\Theta_-^{(2)} Q_-^{(1)} {\Theta'}_+^{(1)}+\Theta_+^{(1)} Q_-^{(1)} {\Theta'}_-^{(2)}+\Theta_-^{(1)} Q_-^{(2)}
   {\Theta'}_+^{(1)}-\Theta_+^{(1)} Q_-^{(2)}
   {\Theta'}_-^{(1)}-\Theta_-^{(1)} Q_+^{(1)} {\Theta'}_-^{(2)}+
\Theta_-^{(2)} Q_+^{(1)} {\Theta'}_-^{(1)}\right]+\nn\\
&-&\Theta_+^{(2)} {Q'}_-^{(1)} Q_-^{(2)} {\Theta'}_+^{(1)}+\Theta_+^{(2)} {Q'}_-^{(1)} Q_+^{(1)} {\Theta'}_-^{(2)}+\Theta_-^{(2)}
   {Q'}_-^{(1)} Q_+^{(2)} {\Theta'}_+^{(1)}-\Theta_+^{(1)}
   {Q'}_-^{(1)} Q_+^{(2)} {\Theta'}_-^{(2)}+
\Theta_+^{(2)} {Q'}_-^{(2)} Q_-^{(1)} {\Theta'}_+^{(1)}+\nn\\
&-&\Theta_+^{(2)} {Q'}_-^{(2)} Q_+^{(1)}
{\Theta'}_-^{(1)}-\Theta_-^{(1)} {Q'}_-^{(2)} Q_+^{(2)}
{\Theta'}_+^{(1)}+ \Theta_+^{(1)} {Q'}_-^{(2)} Q_+^{(2)} {\Theta'}_-^{(1)}\label{defWdelta}
\end{eqnarray}
\end{widetext}
is a generalized Wronskian, it is constant by virtue of the homogeneous system, and therefore can be factored out of the integrals
in Eqs.~\eqref{solQ} and \eqref{solTheta}. In the equations above a prime denotes derivative with respect to the tortoise coordinate $r_*$.

\bibliographystyle{h-physrev4}
\bibliography{dcs_emri_biblio}

\end{document}